\documentclass[%
 reprint,
superscriptaddress,
 amsmath,amssymb,
 aps,
]{revtex4-2}

\usepackage{graphicx}
\usepackage{dcolumn}
\usepackage{bm}

\usepackage{amsmath,amssymb,mathtools}
\usepackage{booktabs,tabularx,array}
\usepackage{xcolor}
\usepackage[hidelinks]{hyperref}

\let\originaleqref\eqref
\renewcommand{\eqref}[1]{Eq.~\originaleqref{#1}}

\newcommand{\E}{\mathbb{E}}
\newcommand{\1}{\mathbf{1}}
\newcommand{\geff}{\Gamma_{\mathrm{eff}}}
\newcommand{\eps}{\varepsilon}
\newcommand{\sw}{\mathrm{sw}}

\newcommand{\Meq}[1]{Eq.~(\ref{#1}) of the main text}
\newcommand{\Meqs}[1]{Eq.~(\ref{#1})}
\newcommand{\Mfig}[1]{Fig.~\ref{#1} of the main text}

\begin{document}

\title{Modeling the Coevolution of Human Skills and AI Adoption}
\author{Giacomo Frigerio}
\affiliation{Dipartimento di Fisica ed Astronomia, Università di Catania and INFN, Catania I-95123, Italy} 
\author{Susanna Caruso}
\affiliation{Dipartimento di Fisica ed Astronomia, Università di Catania and INFN, Catania I-95123, Italy}
\author{Giacomo Livan}
\email[Corresponding author]{ giacomo.livan@unipv.it}
\affiliation{Dipartimento di Fisica, Università di Pavia, Via Bassi 6, 27100 Pavia, Italy}
\affiliation{INFN, Sezione di Pavia, Via Bassi 6, 27100 Pavia, Italy}
\affiliation{Department of Computer Science, University College London, 66-72 Gower Street, WC1E 6EA London, United Kingdom }
\author{Vito Latora}
\affiliation{Dipartimento di Fisica ed Astronomia, Università di Catania and INFN, Catania I-95123, Italy}
\affiliation{School of Mathematical Sciences, Queen Mary University of London, London E1 4NS, United Kingdom}
\affiliation{Complexity Science Hub Vienna, A-1080 Vienna, Austria}

\date{\today}

\begin{abstract}
The value of relying on generative AI depends on two quantities that evolve with its use: users’ unaided skill, which may decline without practice, and the originality of AI-assisted work, which may decline as more people rely on similar tools. 
We model this process as a coevolutionary game in which individuals choose between autonomous and AI-assisted work, with autonomous practice strengthening their skills while reliance on AI gradually erodes them.
When the reinforcing effect of skill erosion outweighs the crowding cost of lost originality, the population exhibits two alternative stable states: low AI adoption with high average skill, and high AI adoption with low average skill. 
Because skill is individual, autonomous workers remain more skilled than AI users, 
making autonomous work less vulnerable to population-wide skill decline.
The strength of this effect is governed by a timescale ratio: the rate at which skills adapt relative to the rate at which individuals change strategy. 
The same ratio determines the full distribution of skills, which we derive analytically. Populations with identical AI adoption and average skill can differ sharply: when strategies change frequently, skills remain homogeneous and collective lock-in to high AI adoption is stronger; when strategies persist, lock-in weakens but the population can polarize into highly skilled autonomous workers and AI-dependent individuals. 
Modeling output overlap further shows that more diverse AI outputs can make widespread adoption harder to reverse. Together, these results expose tradeoffs between the benefits of widespread AI adoption and its long-run risks, making collective lock-in and skill polarization distinct intervention targets.
\end{abstract}

\maketitle

The consequences of generative artificial intelligence (AI) depend not only on what people can accomplish with AI today, but also on how using it changes what they will be able to accomplish without it tomorrow. Generative AI can improve the quality and productivity of creative and knowledge-intensive work~\cite{lee_empirical_2024,boussioux_crowdless_2024,brynjolfsson_generative_2025}, but it may also displace the practice through which individuals develop and maintain their autonomous skills~\cite{wu_how_2026, liu_ai_2026, bastani_generative_2025}. This raises an unresolved question at the heart of the ongoing debate about AI, i.e., whether its short-term benefits may result in the ever increasing dependence on the technology in the long-term~\cite{cash_is_2026}. At the same time, widespread reliance on similar AI systems may make individual outputs increasingly alike~\cite{doshi_generative_2024, meincke_chatgpt_2025, moon_homogenizing_2025},
potentially undermining the collective diversity of ideas even as individual productivity improves. 
Widespread AI adoption may also affect individuals unevenly, deepening skill inequalities and dividing the population into
those who retain their ability to work autonomously and those who become increasingly dependent on AI~\cite{valcea_exploring_2024}. 
With AI capabilities and patterns of use evolving rapidly, understanding how these interconnected processes might unfold requires theoretical models that can explore alternative long-term scenarios and uncover unintended collective consequences. Here, we develop such a model by studying the coevolution of AI adoption and individual skills in a population.

Existing experimental studies provide some initial insights into these concerns and tradeoffs, although most evidence concerns relatively short-term effects. In a short-story task, Doshi and Hauser found that access to generative AI improved evaluations of individual stories, particularly for writers with lower baseline creativity, while making assisted stories more similar to one another \cite{doshi_generative_2024}. Individual gains can thus coexist with a reduction in the collective diversity of outputs. A separate set of randomized experiments by Liu et al. found that AI assistance in mathematical reasoning and reading comprehension improved assisted performance but could reduce subsequent unaided performance \cite{liu_ai_2026}.

These findings, however, do not establish whether the short-term effects of AI assistance translate into lasting skill losses. Moreover, the consequences of AI adoption depend on how assistance is provided and whether users remain actively engaged in the underlying task \cite{moller_overreliance_2026}. In controlled decision-making experiments, AI explanations that required participants to reach their own decisions supported learning, whereas providing recommendations improved immediate decisions without comparable learning gains \cite{gajos_people_2022}. Related studies suggest that the design of AI assistance can either preserve or displace opportunities for independent practice \cite{lira_coach_2026,bastani_generative_2025}.

Understanding how these individual-level effects translate into long-term collective outcomes requires accounting for how people decide whether to use AI. Such decisions may depend on the perceived returns of autonomous and AI-assisted work, including the performance of others, and can give rise to two opposing feedback mechanisms. On the one hand, reliance on AI may displace independent practice, weakening autonomous skills and making further AI adoption increasingly attractive. On the other hand, as AI-assisted outputs become more alike, competition for originality may reduce the returns to AI adoption and discourage further use \cite{moon_homogenizing_2025,azad_ex_2026}. The interplay between these feedbacks, together with differences in individuals' histories of AI adoption, can shape both collective adoption patterns and the distribution of skills across the population.

Theoretical models offer a way to investigate these processes beyond the time horizons accessible to current experiments. Existing approaches have explored several relevant mechanisms, typically focusing on specific aspects of AI adoption and its consequences. At the individual level, models of cognitive offloading examine decisions to delegate work to AI when skills accumulate with practice and decay without it, showing that sufficiently substitutive AI can generate differences in skills across workers \cite{caosun_augmentation_2026,afrouzi_automation_2026}.
Related work has identified tipping points and hysteresis in the coevolution of individual competence and tool reliance, and examined collective lock-in through a population game of outsourcing versus internalization \cite{krakauer_competitive_2026}.
At the population level, models have examined adoption and social welfare when widespread AI adoption affects the performance of AI systems \cite{baumann_generative_2026}, substitution and complementarity between humans and AI through cultural evolution and group selection \cite{zhong_group_2026}, and the diffusion of large language models through social transmission, including tipping points, hysteresis, and lock-in \cite{sole_large-language_2026}. Other work has connected AI-induced crowding of ideas to incentives for adoption \cite{azad_ex_2026}. Together, these studies provide insights into different parts of the problem, while leaving their interplay less well understood.

What remains missing, to the best of our knowledge, is a unified framework in which AI adoption responds to the observed returns of alternative strategies, widespread use generates competition for originality, and individual skills evolve according to each person's history of practice. Importantly, retaining this individual dimension, rather than describing skills solely through population averages, allows us to investigate not only how AI adoption evolves collectively, but also how differences in skills emerge and develop across the population.

To that end, we develop a coevolutionary game in which individuals choose between autonomous and AI-assisted work, while their skills evolve through practice or disuse, and widespread AI adoption reduces the value of originality. Coevolutionary games couple strategies to dynamical variables that they themselves modify~\cite{perc_coevolutionary_2010}. Previous studies have considered individual attributes such as teaching activity~\cite{szolnoki_coevolution_2008}, age~\cite{szolnoki_impact_2009}, and reputation-based partnerships~\cite{fu_reputation-based_2008}. In our model, the coevolving attribute is instead individual skill, whose distribution we characterize analytically.

We find that skill erosion can generate bistability and lock-in to widespread AI adoption. However, differences in individual skills can weaken this tendency, while potentially polarizing the population into highly skilled and AI-dependent individuals. The balance between these outcomes is governed by the relative timescales of skill adaptation and strategy revision. We also show that increasing the diversity of AI outputs can make widespread adoption harder to reverse. These findings highlight distinct, potentially competing objectives for interventions addressing the long-term consequences of AI adoption.

\section{Coevolutionary game of skills and AI adoption}
We consider a population of $N$ agents, each of whom decides whether to rely or not on AI to solve a given task, such as writing a text, generating computer code, producing a visual design, solving a quantitative problem, performing a repetitive task \cite{srivastava_agentic_2026} 
or making a professional judgment.
At time $t$, each agent $i$ is characterized by an individual skill $s_i(t) \in [\eta,1]$, denoting its ability to perform the task without AI assistance. The parameter $\eta \in [0,1]$ sets the lowest possible skill level. 

Each agent can decide between two strategies: AI assisted work (A) or autonomous human work (H). We denote the strategy of each agent ($i$) by $z_i(t) \in \{ H,A \}$. 

Agents adopting strategy $H$ perform the task autonomously. Their performance thus depends directly on the current individual skill, and we assign them a payoff:  
\begin{equation}
    \label{eq:payoffH}
    \pi_H = b_h s_i, 
\end{equation}
where the parameter $b_h>0$ sets the benefit associated with this choice. 

Agents adopting strategy $A$, instead, rely on AI assistance. We assume that AI adoption provides a baseline benefit $b_a>0$, while its effectiveness decreases as adoption becomes more widespread in the population. 
We thus define:
\begin{equation}
   \label{eq:payoffA} 
   \pi_A = b_a - c x^{\alpha} \ ,
\end{equation}
where $x(t)$ is the fraction of agents adopting AI at time $t$.
The parameter $c>0$ controls the strength of the crowding effect, while the synergistic exponent 
$\alpha$ governs the power-law dependence on the fraction $x$ of AI users. 

The crowding term reflects the fact that in many settings the value of an output depends not only on its quality but also on its originality. Creativity is commonly defined as the combination of novelty and usefulness \cite{runco_standard_2012}; scientific credit rewards priority \cite{merton_priorities_1957}; and content that competes for limited attention loses visibility when similar contents are available \cite{weng_competition_2012}.
An output that resembles many competing ones is therefore worth less, even when produced competently. Outputs produced with generative AI are measurably more similar to one another than those produced without it \cite{doshi_generative_2024,meincke_chatgpt_2025,moon_homogenizing_2025}, so the crowding penalty grows with the fraction of AI users. When few people use AI, AI-assisted work remains distinctive, and its value declines as adoption spreads \cite{azad_ex_2026}. The cost is thus imposed by similarity to others, not by the use of AI as such. 
At the end of this section, we trace its origin from the overlap between human and AI-assisted outputs
and discuss the effects of the synergistic exponent $\alpha>1$ and $\alpha<1$.

Agents revise their strategy through stochastic pairwise comparison \cite{traulsen_stochastic_2006,traulsen_pairwise_2007}. 
A focal agent $i$ samples another agent $j$ and adopts their strategy with probability
\begin{equation}
    P(i\leftarrow j)
    =
    \frac{1}{1+\exp[-\beta(\pi_j-\pi_i)]},
    \label{eq:fermi}
\end{equation}
where $\beta$ controls the strength of payoff-based selection. 
This rule encodes social learning from observed performance, capturing the possible unintended consequences of AI adoption. Indeed, an agent's payoff reflects both their individual skill and their current strategy. Thus, the imitation of strategies adopted by high-payoff-agents via~\eqref{eq:fermi} may disadvantage the imitating agent, as such a process is blind to individual skill \cite{livan_dont_2019,de_domenico_imitation_2024}.

In addition to payoff-driven imitation, we allow for spontaneous strategy changes occurring at rate $\mu$. These changes are independent of payoff differences and represent exploration, experimentation or idiosyncratic changes in behavior. This parameter therefore introduces a persistent source of behavioral variability.    

Strategy choice not only determines the immediate payoff obtained by an agent, but also affects the evolution of its ability to perform the task autonomously. We therefore let individual skill adapt over time according to the strategy currently adopted. We assume
\begin{equation}
    \dot{s}_i=
    \begin{cases}
        \varepsilon(1-s_i), & z_i=H,\\
        \varepsilon(\eta-s_i), & z_i=A.
    \end{cases} \label{eq:si-evolution}
\end{equation}
where $\varepsilon$ controls the characteristic timescale of skill adaptation. Previous studies on skill decay under disuse imply half-lives of weeks to months for cognitive skills  \cite{arthur_jr_factors_1998,krakauer_competitive_2026}, which sets the real-time scale of $\varepsilon$. \eqref{eq:si-evolution} has the learning-forgetting structure used in recent models of AI-induced skill atrophy \cite{ganuthula_paradox_2026,krakauer_competitive_2026,caosun_augmentation_2026}.
Thus, autonomous work reinforces the agent's skill and drives $s_i$ to its maximal value $1$, whereas
the adoption of AI progressively reduces autonomous skill toward the lowest level $\eta$. 
Since agents may switch strategy over time, their current skill depends on the history of each agent's strategy adoption.
Strategy choices $\{ z_i \}$ and skills $\{ s_i \}$ therefore dynamically coevolve: the strategy currently adopted determines the evolution of an agent's skill, where skills in turn affect the payoff associated with autonomous work and hence the relative attractiveness of the two strategies.  
Our microscopic agent model keeps track of the strategy and skill of every individual agent and therefore allows, in principle, for heterogeneous skills and correlations between strategy and skill.

The model belongs to the class of coevolutionary games, in which agents' strategies coevolve with additional dynamical variables describing, for instance, the state of the environment \cite{tilman_evolutionary_2020,hauert_asymmetric_2019,weitz_oscillating_2016,su_evolutionary_2019} or other relevant features of the system \cite{perc_coevolutionary_2010,pacheco_coevolution_2006}.
In this literature, strategies have been coupled with individual attributes, such as teaching activity \cite{szolnoki_coevolution_2008,szolnoki_cooperation_2007}, aging \cite{szolnoki_impact_2009}, reputation \cite{fu_reputation-based_2008}. 

In our model, the coevolving attribute is individual skill, which is specific to each agent and directly affects the payoff of autonomous work. This allows differences in skill levels to emerge across agents and may create correlations between strategy and skill even in an initially homogeneous population. We next derive a mean-field description of this feedback and examine how it shapes the collective dynamics.

\medskip
\section{Mean-field theory}

In order to obtain analytical insights on the collective dynamics of our coevolving model of human skills and AI adoption, we derive a macroscopic description of the well-mixed population in terms of AI adoption and mean skill  
in the large-population and weak-selection limit (see Supplemental Material (SI)). The fraction of AI users is
\begin{equation}
    x(t)=\frac{1}{N}\sum_{i=1}^{N}\mathbf{1}(z_i(t)=A),
\end{equation}
where $x=0$ and $x=1$ correspond, respectively, to the absence and universal adoption of AI in the population. In this mean-field limit, the dynamics can be expressed in terms of a smaller number of macroscopic variables. 
Due to the linearity of~\eqref{eq:si-evolution}, by averaging over agents we obtain an equation for the mean skill $s=N^{-1}\sum_i s_i=(1-x)s_h+xs_a$, where $s_h=\mathbb{E}[s_i\mid z_i=H]$ and $s_a=\mathbb{E}[s_i\mid z_i=A]$ are the mean skills of autonomous and AI-assisted agents:  
\begin{equation}
\dot{s}=\varepsilon\left[1-(1-\eta)x-s\right].
    \label{eq:reduced-s}
\end{equation}
Such an equation expresses the rate of change of $s$ as a function of $s$ itself and of the fraction $x$ of AI adopters.

In order to get an equation for the 
time evolution of $x$, notice that 
the AI payoff $b_a-cx^{\alpha}$ is the same for all agents adopting AI, whereas the payoff $b_h s_i$ of autonomous agents depends on their individual skill. Hence, every $H$--$A$ comparison for strategy selection needs to take into account the skill of the $H$ agent. 
In the large-population and
weak-selection limit, and after rescaling time by the selection intensity (see SI), the fraction of AI users obeys
\begin{equation}
    \dot{x}=x(1-x)\left[b_a-cx^{\alpha}-b_hs_h\right]+m(1-2x),
    \label{eq:replicator-sh}
\end{equation}
where $m$ is the spontaneous switching rate in rescaled units (see SI for its relation to $\mu$). Strategy selection therefore depends on $s_h$ rather than on $s$. Selection compares AI users with the agents currently working autonomously (i.e., choosing strategy $H$), that may be more skilled than the population average. 

\begin{figure*}[t]
    \centering
    \includegraphics[width=0.98\textwidth]{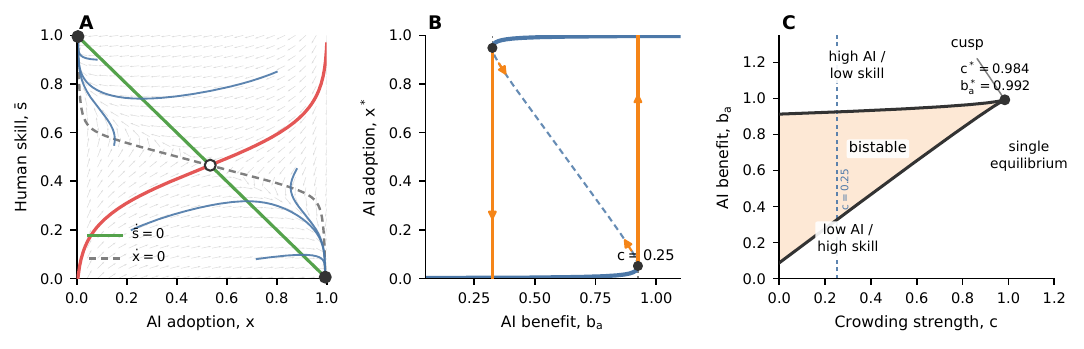} 
    \caption{\textbf{Bistability and hysteresis in the reduced mean-field theory (Reduced MFT).} \textbf{(A)} Phase portrait in the bistable regime ($b_a=0.6$). Green solid and gray dashed curves in the $(x,s)$ plane  
are the nullclines $\dot{s}=0$ and $\dot{x}=0$; filled circles are the low-AI/high-skill and high-AI/low-skill equilibria;  the open circle is the saddle, and the red curve is its stable manifold, which separates the basins. Blue curves are representative trajectories. \textbf{(B)} Equilibrium AI adoption $x^{\star}$ as a function of the AI (baseline) benefit $b_a$. Solid branches are stable and the dashed branch is a saddle. Arrows show the quasi-static hysteresis loop: increasing $b_a$, the population remains on the low-adoption branch up to the fold $b_{a,+}\simeq0.925$ and then jumps to high adoption; decreasing $b_a$, it remains on the high-adoption branch down to $b_{a,-}\simeq0.325$. Mean skill traces the mirror-image loop. \textbf{(C)} Fold points $b_{a,\pm}$ as functions of the crowding strength $c$. The shaded region is bistable and closes at the cusp where $\Gamma=8m$, $c^{*}=b_h(1-\eta)-8m$; for $c>c^{*}$ a single equilibrium exists for every $b_a$. The vertical line marks the value of $c$ used in A and B. Parameters: $b_h=1$, $\eta=0$, $\varepsilon=0.05$, $m=0.002$, $\alpha=1$, $c=0.25$ unless stated.
}
    \label{fig:meanfield}
\end{figure*}

\medskip
\subsection{Reduced mean-field theory}

We first consider a reduced description in which the mean skill of autonomous agents is approximated by the population average, $s_h\simeq s$. This approximation is expected to hold when agents switch strategies much faster than their skills change, limiting the skill differences that develop between autonomous and AI-assisted agents. 
Under this approximation, \eqref{eq:replicator-sh} reduces to 
\begin{align}
    \dot{x}&=x(1-x)\left[b_a-cx^{\alpha}-b_hs\right]+m(1-2x),
    \label{eq:reduced-x}
\end{align}
which, together with \eqref{eq:reduced-s}, forms a closed system for $x$ and $s$, that we will refer to as the reduced mean-field theory (Reduced MFT).
At equilibrium, \eqref{eq:reduced-s} gives:  
\begin{equation}
 s^{\star}=1-(1-\eta)x^{\star}. 
 \label{eq:s-star}
\end{equation}
Since $\eta<1$, higher AI adoption is therefore associated with lower mean skill. 
For linear crowding ($\alpha=1$), the stationary payoff difference can be written as
\begin{equation}
\Delta\pi^\star= \pi^\star_A - \pi^\star_H=
b_a-b_h+\Gamma x^\star, 
\label{eq:delta-diff}
\end{equation}
where $\pi^\star_H =  b_hs^\star$, and
\begin{equation}
    \Gamma=b_h(1-\eta)-c  
    \label{eq:gamma-feedback}
\end{equation}
measures the balance between the skill-mediated reinforcement of AI adoption, $b_h(1-\eta)$, and the opposing effect of crowding, $c$.

To clarify the role of these two terms, consider first the case without spontaneous switching, $m=0$. As AI adoption increases, skill erosion lowers the payoff of autonomous work and therefore makes strategy $A$ (AI adoption) relatively more attractive. Crowding acts in the opposite direction, because a larger fraction of AI users lowers the payoff of strategy $A$ itself. When $\Gamma>0$, the first effect dominates and AI adoption becomes self-reinforcing. 
For a benefit $b_a$ of AI  adoption such that:
$$
b_h\eta+c<b_a<b_h,
$$
both the state $(x,s)=(0,1)$,  
in which all agents adopt strategy 
$H$, and the state $(1,\eta)$, 
in which all agents adopt strategy 
$A$ (i.e. decide to use AI), are stable, and are separated by an interior saddle. When $\Gamma<0$, crowding dominates instead: increasing AI adoption reduces its own relative payoff, and in the interval 
$$
b_h<b_a<b_h\eta+c
$$
the system converges to a stable interior equilibrium in which the two strategies coexist (see SI). Skill erosion alone is therefore not sufficient for bistability; its reinforcing effect must outweigh crowding.

For $m>0$, instead, spontaneous strategy switching removes the absorbing boundary states, although it does not necessarily eliminate the two-attractor structure. 
{In particular, there exists a range of $b_a$ values over which the system is bistable if and only if (see SI)}: 
\begin{equation}
    \Gamma>8m .
    \label{eq:bistability-condition}
\end{equation}
Within this regime, two stable equilibria coexist, each with its own basin of attraction [Fig.~\ref{fig:meanfield}A]. Which equilibrium the population approaches depends on both the initial AI adoption level and the initial mean skill level: populations starting with the same adoption level can reach different long-run states, if their initial skills differ.
The bistable region is bounded by two saddle-node (fold) bifurcations and gives rise to hysteresis when the baseline benefit $b_a$ of AI is varied slowly. As $b_a$ increases, the population remains on the low-adoption branch until the upper fold $b_{a,+}$, where the stable equilibrium collides with the saddle and disappears, forcing the system to jump to the high-adoption branch. When $b_a$ is subsequently decreased, the population remains on the high-adoption branch until the lower fold $b_{a,-}<b_{a,+}$, where the corresponding stable equilibrium disappears and the system jumps back [Fig.~\ref{fig:meanfield}B]. Since mean skill decreases with adoption, as per \eqref{eq:s-star}, it follows the corresponding loop in the opposite direction: skill drops at $b_{a,+}$ and recovers only at $b_{a,-}$. Restoring an earlier value of $b_a$ therefore need not restore the earlier population state.
Finally, increasing crowding reduces $\Gamma$ and progressively narrows the bistable region. The two folds meet at
$$
c^{\star}=b_h(1-\eta)-8m,
$$
beyond which only a single equilibrium remains [Fig.~\ref{fig:meanfield}C].

An analogous hysteresis behavior has been observed in a model of a single user whose reliance and competence coevolve \cite{krakauer_competitive_2026}. In our model, it emerges at the population level from the competition between skill erosion and frequency-dependent crowding. Nonlinear crowding preserves this mechanism while allowing tipping between autonomy and partial AI adoption ($\alpha>1$), or between partial and near-complete AI adoption ($\alpha<1$), even when $\Gamma\leq0$ (SI). Saturating crowding of the kind estimated in \cite{azad_ex_2026} corresponds to the latter.

\medskip
\section{Individual skills: agent-based simulations and mean-field closure}

We now return  to the microscopic model and simulate the coupled evolution of individual skills and strategies. This allows us to identify when the reduced MFT description applies and how emerging skill differences between strategy classes modify bistability. In our 
agent-based model 
we simulate a finite population in which 
each agent $i$ has a skill $s_i$ that evolves in time according to~\eqref{eq:si-evolution}, and updates their strategy according to~\eqref{eq:fermi}. 
Individual skills evolve continuously according to the strategy currently adopted, while strategy changes occur through imitation and spontaneous switching. The agent-based simulations therefore retain the whole distribution of individual skills and any correlations that emerge between strategies and skills. Details of the numerical simulations are provided in Methods and in the SI.

We find that bistability and hysteresis persist in the agent-based model,  
although the resulting hysteresis loop is substantially narrower than that predicted by the reduced MFT. [Fig.~\ref{fig:agent-hysteresis}A,B]. While the latter predicts a bistable interval $0.33<b_a<0.92$ independently of $\varepsilon$, in the numerical simulations at $\varepsilon=2$ the upward and downward transitions appear to be much closer. To understand the origin of this narrowing of the bistability region 
we revisit the approximation $s_h\simeq s$ underlying \eqref{eq:reduced-x}. 

\subsection{Mean-field theory with covariance closure}

In the agent-based model, each individual's skill reflects their strategy history, generating correlations between current strategy and skill that this approximation neglects.
Introducing the strategy--skill covariance: 
\begin{equation}
    \sigma =\mathbb{E}\left[\mathbf{1}(z_i=A)s_i\right]-xs=x(1-x)(s_a-s_h) \ ,
    \label{eq:workflow-skill-covariance}
\end{equation}
and plugging $s_h=s-\sigma/(1-x)$ in \eqref{eq:replicator-sh}, allows 
to rewrite it as a function of $s$, $x$ and $\sigma$:  
\begin{equation}
    \dot{x}=x(1-x)\left[b_a-cx^{\alpha}-b_hs\right]+b_hx\sigma+m(1-2x).
    \label{eq:replicator-covariance}
\end{equation}
When the expected skill of autonomous agents exceeds that of AI users, $\sigma$ becomes negative, and so does the correction term $b_hx\sigma$ above, thereby slowing down AI adoption in~\eqref{eq:replicator-covariance}.
The size of this correction is  controlled by the time evolution 
of the covariance, which depends on how switching responds to individual skills. 
In order to write an equation for the time evolution of $\sigma$,
let $r_{+}$ and $r_{-}$ denote the per-capita rates at which agents switch to AI assistance and AI-assisted agents switch back, so that $\dot{x}=r_{+}(1-x)-r_{-}x$. 
Approximating these rates as independent of the switching agent's own skill gives: 
\begin{equation}
  \dot{\sigma}= - \varepsilon \left[ \left(1+\frac{\lambda}{\varepsilon}\right)\sigma+(1-\eta)\,x(1-x) \right],
  \label{eq:C_dynamics}
\end{equation}
where we have set 
$\lambda=r_{+}+r_{-}$. 
\eqref{eq:C_dynamics} is exact for skill-independent switching. In the payoff-driven model, skill-dependent switching adds a conditional covariance term, derived in the SI. Finally, 
 \eqref{eq:C_dynamics}, \eqref{eq:reduced-s} and \eqref{eq:replicator-covariance} close the mean-field equations at the level of the strategy--skill covariance, and govern the evolution of the three state variables: the fraction $x$ of agents with strategy $A$, the population mean skill $s$,  and the strategy--skill covariance $\sigma$.
We refer to them as the mean-field theory (MFT) with covariance closure.

Notice that the second term in \eqref{eq:C_dynamics} is negative whenever $0<x<1$ and $0\le\eta<1$. Hence, agents currently working autonomously become, on average, more skilled than those relying on AI, even in the absence of initial correlations between skills and strategies.
This has two closely related consequences: individuals adopting different strategies develop systematically different skill levels, and the resulting skill gap alters the feedback mechanism underlying bistability.
Both effects are controlled by the relative timescales of skill adaptation and strategy switching. We define
\begin{equation}
    \rho=\frac{\varepsilon}{\lambda},
    \label{eq:rho}
\end{equation}
where $\varepsilon$ is the rate of skill adaptation and $\lambda=r_++r_-$ is the decorrelation rate of the strategy process.
Thus, $1/\varepsilon$ is the characteristic time over which skills adjust, whereas $1/\lambda$ is the characteristic time over which an agent's current strategy ceases to predict its future strategy. The ratio $\rho$ therefore compares how quickly skills respond to a strategy with how long that strategy persists.

When $\rho\ll1$, agents switch strategies many times before their skills can adapt appreciably. Individual skill therefore reflects a mixture of recent strategy choices, and agents currently choosing strategy $H$ are only slightly more skilled than the population as a whole. In this regime, strategy--skill correlations are weak and the closure $\sigma\simeq0$ provides a good approximation. By contrast, when $\rho\gtrsim1$, agents remain with the same strategy long enough for their skills to approach the corresponding target level. Agents choosing $H$ thus maintain systematically higher skills than those choosing $A$, generating a negative strategy--skill covariance and a persistent skill gap between the two groups. At stationarity, \eqref{eq:C_dynamics} gives
\begin{equation}
  \sigma^{\star}=-(1-\eta)\,x^{\star}(1-x^{\star})\,\frac{\rho}{1+\rho},
  \qquad
  s_h^{\star}-s_a^{\star}=(1-\eta)\,\frac{\rho}{1+\rho}.
  \label{eq:C_stationary}
\end{equation}
which shows that coevolution alone generates systematic heterogeneity between strategy classes.
Even when agents start with the same skills and their skills respond in the same way to practice and disuse, persistent strategy choices sort them into groups with different skill levels: autonomous workers become systematically more skilled than AI users. 
Skill heterogeneity therefore emerges endogenously from the coupling between strategy and skills, rather than having to be assumed as a pre-existing difference between individuals.

\eqref{eq:C_stationary} is exact for skill-independent switching. The gap vanishes when strategy switching is fast relative to skill adaptation ($\rho\to0$), and approaches its maximal value $1-\eta$ when strategies persist long enough for each group to relax toward its strategy-specific skill level ($\rho\to\infty$). The agent-based model confirms this prediction directly: the normalized skill gap $(s_h-s_a)/(1-\eta)$ follows $\rho/(1+\rho)$ over two decades in $\rho$, with no fitted parameters [Fig.~\ref{fig:agent-hysteresis}C]. 
More generally, an identity that follows from the stationarity of the second moment of skill, and therefore holds for any switching rule (see SI), links the gap to the variance of individual skills:
\begin{equation}
   s_h^\star-s_a^\star=\frac{\operatorname{Var}(s)}{(1-\eta)\,x^\star(1-x^\star)}.
\end{equation}
At a given adoption level, a wider skill distribution therefore implies a larger gap between strategy classes. 
Since autonomous skills are estimated to decay over weeks to months under disuse \cite{krakauer_competitive_2026}, $\rho$ is expected to be small when individuals reconsider their use of AI on a timescale of days, and of order one or larger when strategy choices persist for months. Importantly, $\rho$ is not an additional free parameter: it follows from the microscopic update rule and, in the weak-selection regime considered here, is approximately proportional to $\varepsilon$ (see SI).

\subsection{Skill gap narrows the bistability region.}
\begin{figure*}[t!]
    \centering
    \includegraphics[width=.75\textwidth]{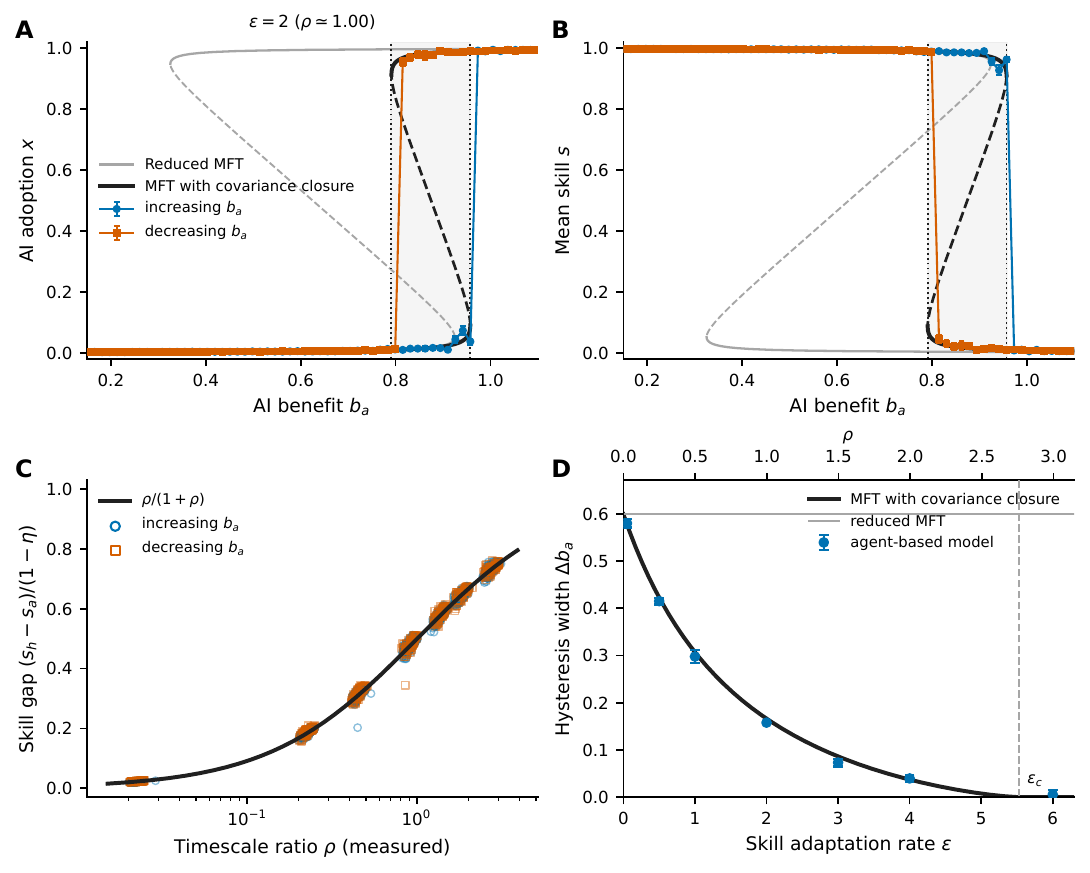}
    \caption{\textbf{Individual skills generate skill gaps and narrow the hysteresis loop.}
    \textbf{(A)} Fraction $x$ of AI users and \textbf{(B)} mean skill $s$ as a function of the benefit $b_a$ of AI adoption. 
    Skills adapt at $\varepsilon=2$ ($\rho\simeq1.0$).
    Symbols denote the results of the agent-based model under a quasi-static sweep of $b_a$, increased from $0.15$ to $1.10$ (blue circles) and then decreased over the same interval (orange squares). 
Error bars are temporal standard deviations over the measurement window.
Gray curves report the stationary state predicted  by the reduced MFT ($\sigma=0$ and $\Gamma$ as in \eqref{eq:gamma-feedback}),  
while black curves are those predicted by the MFT with covariance closure
($\Gamma_{\mathrm{eff}}$ as in \eqref{eq:gamma-eff}). Solid lines and dashed lines denote stable and unstable branches, respectively.  
The shaded intervals mark the bistable region predicted by the effective-feedback $\Gamma_{eff}$, bounded by the folds at $b_{a,-}\simeq0.79$ and $b_{a,+}\simeq0.96$.
    \textbf{(C)} Normalized skill gap between autonomous and AI-assisted agents, $(s_h-s_a)/(1-\eta)$, against the timescale ratio $\rho=\varepsilon/\lambda$, with $\lambda$ the total switching rate measured in each stationary state. Points pool all values of $\varepsilon$, both sweep directions, and all repetitions; states in which either strategy is used by fewer than $10$ agents are excluded. Black line: $\rho/(1+\rho)$ (\eqref{eq:C_stationary}), with no fitted parameters.
    \textbf{(D)} Width of the hysteresis loop, $\Delta b_a$, against the skill  adaptation rate $\varepsilon$ (top axis: predicted $\rho$). Black curve: MFT with covariance closure, with the loop closing at $\varepsilon_c\simeq5.53$ ($\rho_c\simeq2.76$, dashed line). Gray line: width predicted by the reduced MFT  with $\sigma=0$, independent of $\varepsilon$. Points: agent-based model, mean and standard deviation over $3$ repetitions. The simulated width is the difference between the values of $b_a$ at which the increasing and decreasing branches cross $x=1/2$, obtained by linear interpolation. 
    Parameters: $b_h=1$, $c=0.25$, $\eta=0$, $\alpha=1$, $m=0.002$, $\beta=0.5$; $N=4000$ agents; crowding estimated from $k=50$ agents sampled uniformly at each payoff evaluation. The sweep uses $61$ equally spaced values of $b_a$; at each value the population relaxes for $1200$ sweeps, and measurements are taken every $5$ sweeps over $300$ further sweeps. Simulations are event-driven: each agent revises its strategy at unit rate, exploring with probability $\mu\simeq5.0\times10^{-4}$ per revision (corresponding to $m=0.002$), and skills evolve exactly between revisions. $\varepsilon$ and $m$ are expressed in rescaled time, obtained from sweeps by the factor $(1-\mu)\beta/2$. Values of $\varepsilon$ in \textbf{C} and \textbf{D}: $0.05$, $0.5$, $1$, $2$, $3$, $4$ and $6$. The upward sweep starts from $x=0.02$ with $s_i=1$ for all agents; the downward sweep starts from the final state of the upward one.}
    \label{fig:agent-hysteresis}
\end{figure*}

The skill gap in turn changes the collective dynamics. The payoff of strategy $H$ depends on the skill of agents currently choosing $H$, rather than on the population-average skill. Combining the stationary mean skill, \eqref{eq:s-star}, with \eqref{eq:C_stationary} through $s_h=s+x(s_h-s_a)$ gives
\begin{equation}
    s_h^\star
    =
    1-\frac{(1-\eta)x^\star}{1+\rho}.
    \label{eq:sh-star}
\end{equation}
The stationary payoff difference, for linear crowding $\alpha =1$, therefore retains the same form as
in the reduced MFT (see \eqref{eq:delta-diff}):
\begin{equation}
    \Delta\pi^\star
    =
    b_a-b_h+\Gamma_{\mathrm{eff}}x^\star,
\end{equation}
where:
\begin{equation}
    \Gamma_{\mathrm{eff}}
    =
    \frac{b_h(1-\eta)}{1+\rho}-c .
    \label{eq:gamma-eff}
\end{equation}
The skill-mediated reinforcement of AI adoption is therefore reduced by the factor $1/(1+\rho)$. As the skill gap widens, agents currently working autonomously remain more skilled than the population average and are partly shielded from the overall decline in skill. Consequently, as AI adoption increases, the payoff of strategy $H$ declines more slowly than predicted by the reduced MFT.
This provides a direct link between the individual skill gap and collective bistability. As $\rho$ increases, $\Gamma_{\mathrm{eff}}$ decreases and the hysteresis region narrows.
Within the approximation that $\rho$ remains constant along the equilibrium branch, the stationary equation has the same form as in the reduced MFT, with $\Gamma$ replaced by $\Gamma_{\mathrm{eff}}$.
For linear crowding, a bistable interval exists for some values of $b_a$ if and only if
\begin{equation}
    \frac{b_h(1-\eta)}{1+\rho}-c>8m.
\end{equation}
When $\rho$ varies with adoption, the turning points must instead be obtained by differentiating the full equilibrium curve, including this dependence. The resulting condition is given in the SI.
For $\rho\to0$, the skill gap vanishes and $\Gamma_{\mathrm{eff}}\to\Gamma$, recovering the 
results of the 
reduced MFT.  Conversely, when skill adaptation is fast relative to strategy switching, autonomous agents retain high skill even at substantial levels of AI adoption, progressively weakening the reinforcing loop between AI adoption and skill loss until bistability disappears.
The predictions of the MFT with covariance closure are in good agreement with the simulations of the agent-based model. 
The theory based on $\Gamma_{\mathrm{eff}}$ predicts $b_{a,-}\simeq0.79$ and $b_{a,+}\simeq0.96$, capturing the transitions of the simulated hysteresis loop [Fig.~\ref{fig:agent-hysteresis}A,B]. The width of the hysteresis loop decreases as skill adaptation becomes faster relative to strategy switching and vanishes at $\varepsilon_c\simeq5.5$, in agreement with the prediction based on $\Gamma_{\mathrm{eff}}$ [Fig.~\ref{fig:agent-hysteresis}D]. The critical value follows from the loss of bistability, $\Gamma_{\mathrm{eff}}=8m$, which gives
$$\rho_c=\frac{b_h(1-\eta)}{c+8m}-1, $$
and, using $\rho=\varepsilon/\lambda$, yields $\varepsilon_c\simeq5.5$ for the parameters of Fig.~\ref{fig:agent-hysteresis}.

\medskip
\subsection{Distribution of individual skills}
\begin{figure*}[t] 
\centering
\includegraphics[width=\textwidth]{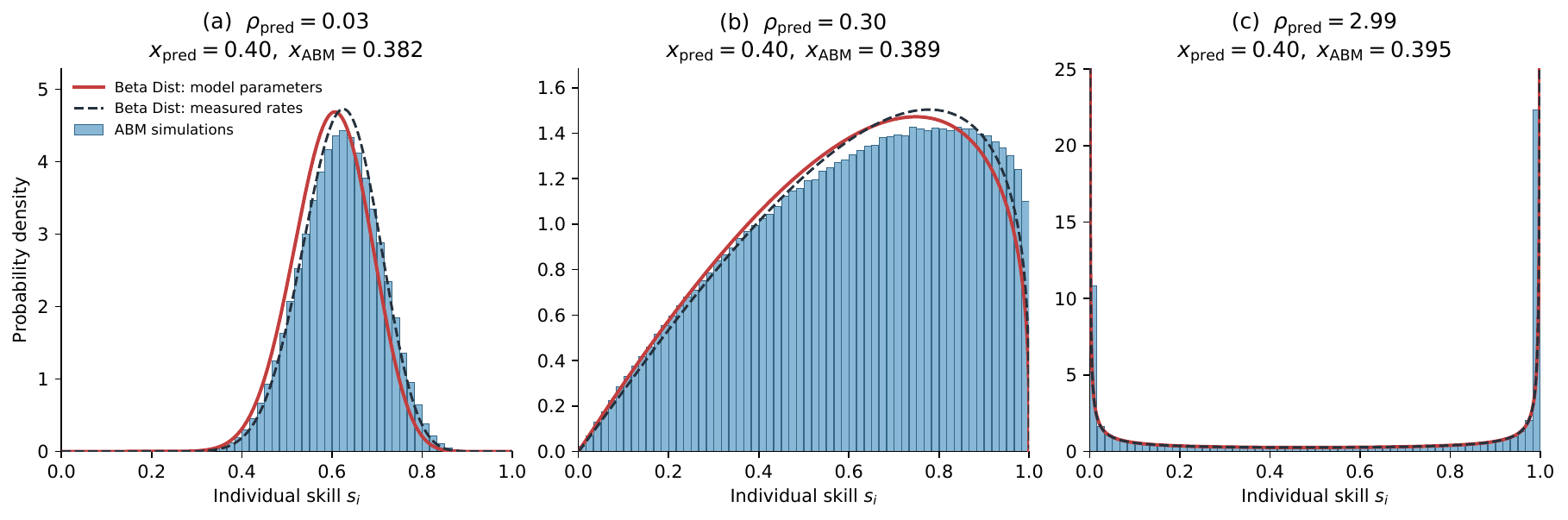}
\caption{\textbf{Distribution of individual skills.}
Stationary distribution of individual skill $s_i$ in the agent-based model (histograms) for three values of the timescale ratio $\rho$.
The solid red curve is the Beta distribution of \eqref{eq:beta-law}, with $x^{\star}$ and $\rho$ predicted entirely from the model parameters; the dashed curve shows the corresponding Beta law obtained using the switching rates measured directly in the simulations. \textbf{(A)} $\rho_{\rm pred}\simeq0.03$: strategy switching is fast relative to skill adaptation, and individual skills remain concentrated around the population mean. \textbf{(B)} $\rho_{\rm pred}\simeq0.30$: switching and skill adaptation occur on comparable timescales, producing a broad and asymmetric distribution over the skill interval. \textbf{(C)} $\rho_{\rm pred}\simeq2.99$: skill adaptation is fast relative to strategy switching, and the distribution becomes U-shaped, with probability concentrated near the strategy-specific attractors $s_i=\eta$ and $s_i=1$.
In all three panels the theoretical equilibrium is fixed at $x^{\star}=0.40$, while the corresponding agent-based values are $x_{\rm ABM}=0.382$, $0.389$, and $0.395$, respectively. For $\eta=0$, the theory therefore predicts the same population-average skill, $s^{*}=1-x^{\star}=0.60$, in all panels, illustrating that similar mean skill can coexist with qualitatively different underlying skill distributions.
Parameters are $b_h=1$, $c=1.2$, $\eta=0$, selection intensity $\beta=1$, $m=0.002$, $N=3000$, and $k=50$. The AI payoff intercept is chosen separately in each panel so that the theory predicts $x^{\star}=0.40$, giving $b_a=1.090$, $1.170$, and $1.378$ for panels \textbf{A--C}, respectively.
The manuscript values $\varepsilon=0.03$, $0.30$, and $3.0$ correspond to per-sweep relaxation rates $\varepsilon_{\rm sweep}=0.01499$, $0.14985$, and $1.49850$ under the time convention used in the simulations. Each agent undergoes stochastic Fermi revision events at unit rate; exploration occurs with per-revision probability $\mu\simeq9.99\times10^{-4}$, corresponding to $m=0.002$ in manuscript units.
Crowding is estimated from $k=50$ sampled agents. Runs start from $x(0)=0.40$ with randomly assigned strategies and $s_i(0)=1$ for all agents. Skill relaxation is evolved exactly between revision events. After a burn-in of approximately $1668$ sweeps in panel \textbf{A} and $600$ sweeps in panels \textbf{B--C}, distributions are accumulated over a further $1800$ sweeps, using one snapshot every $5$ sweeps ($360$ snapshots) and $60$ histogram bins.}
    \label{fig:agent_capacity_distributions}
\end{figure*}

The agent-based model produces a full distribution of individual skills, shaped by each agent's history of switching between strategies. 
The shape of such distribution varies with the timescale ratio $\rho$ 
[Fig.~\ref{fig:agent_capacity_distributions}], 
and can be understood through the following argument. 
Each agent alternates between two regimes: while not aided by AI, their skill rises toward $1$; while relying on AI, it decays toward $\eta$ (as per \eqref{eq:si-evolution}). Switches between the two regimes occur at random times, with rates set endogenously by the population. If the switching rates did not depend on individual skill, this process would be exactly the one studied for variables driven by dichotomous noise and for two-state models of gene expression, whose stationary distribution is a Beta distribution \cite{peccoud_markovian_1995,werner_horsthemke_noise-induced_1984}. 
We use this distribution as an analytical approximation to the 
case of 
the agent-based
model, in which switching is payoff-driven and depends on individual skills. 
Defining the normalized skill 
$u=(s-\eta)/(1-\eta)$, the approximation gives (SI Appendix)
\begin{equation}
    u\sim\mathrm{Beta}\!\left(\frac{1-x^{\star}}{\rho},\,\frac{x^{\star}}{\rho}\right),
    \label{eq:beta-law}
\end{equation}
where $x^{\star}$ is the stationary fraction of AI users  as predicted by $\Gamma_{\mathrm{eff}}$, and $\rho$ follows from the update rule. 

Although the Beta distribution is approximate, its mean, $\langle u\rangle=1-x^\star$, agrees with the exact stationary relation obtained from the mean-skill dynamics in \eqref{eq:reduced-s}. At fixed $x^\star$, the shape of the distribution is controlled by $\rho$.
For $\rho\ll1$ individual histories average out and skills concentrate around the mean. For intermediate $\rho$ the distribution spreads over most of the skill interval. For $\rho\gtrsim1$ the distribution of skills becomes U-shaped, with agents accumulating near full skill and near the lowest possible skill level $\eta$. This change of shape is the analog of the noise-induced transition known for dichotomous noise \cite{sancho_stochastic_1984,werner_horsthemke_noise-induced_1984}. 
The Beta distribution in \eqref{eq:beta-law} computed using the model parameters (red curves) nearly coincides with the one obtained from the switching rates measured in the simulations (dashed blue curves), and both reproduce well the results of simulated distributions [Fig.~\ref{fig:agent_capacity_distributions}]. 
The main deviation, an excess of highly skilled agents at intermediate $\rho$ [Fig.~\ref{fig:agent_capacity_distributions}B], has a simple origin: highly skilled autonomous agents earn higher payoffs and are therefore less likely to adopt AI, remaining autonomous for longer than 
predicted by an approximation in which strategy switching is independent of individual skill.

These distributions show that the population-average skill does not fully characterize the state of the system. At fixed $x^\star$, the exact mean-skill relation fixes the same average skill independently of $\rho$, while \eqref{eq:beta-law} predicts qualitatively different distributions around that mean. A population in which most agents have intermediate skill can therefore have the same AI adoption and mean skill as one divided between highly skilled and strongly deskilled individuals. The timescale ratio $\rho$ thus controls how skill is distributed across the population even when aggregate quantities remain unchanged.
Taken together with the results of Fig.~\ref{fig:agent-hysteresis}, this reveals a tradeoff between collective history dependence and skill polarization. When $\rho\ll1$, frequent strategy switching keeps individual skills concentrated around the mean, but strategy--skill correlations are weak and the reinforcing feedback driving bistability remains close to its full strength. As $\rho$ increases, persistent strategy choices generate a larger skill gap and weaken this collective feedback, narrowing the hysteresis region, while at the same time broadening the distribution of individual skills and eventually producing a polarized population. Collective lock-in and skill polarization are therefore distinct outcomes: reducing one does not necessarily reduce the other.

\begin{figure*}[t]
    \centering
    \includegraphics[width=.98\textwidth]{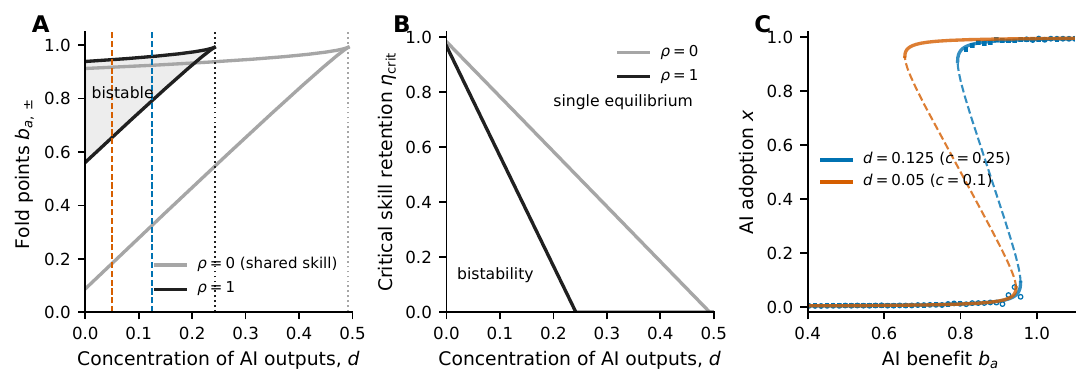}
    \caption{\textbf{More diverse AI output makes widespread AI adoption harder to reverse.} 
    Predictions of the MFT with covariance closure and    
    with crowding generated by output overlap, $c=\omega d$ [\eqref{eq:c-from-d}], where $d$ measures the excess concentration of AI-assisted outputs relative to the autonomous benchmark. \textbf{(A)} Saddle-node (fold) points $b_{a,\pm}$ as functions of $d$ for the Reduced MFT
        ($\rho=0$, gray) and with strategy--skill correlations ($\rho=1$, black). The region between the two folds is bistable. Reducing $d$ lowers the lower fold substantially while barely moving the upper fold. Dotted vertical lines mark the critical concentration $d_c$ [\eqref{eq:d-critical}] for the two values of $\rho$; colored dashed lines mark the two values of $d$ shown in C. \textbf{(B)} Critical skill retention $\eta_{\mathrm{crit}}$ above which bistability disappears, leaving a single equilibrium for every $b_a$, as a function of $d$ for $\rho=0$ (gray) and $\rho=1$ (black). \textbf{(C)} Equilibrium AI adoption as a function of $b_a$ at $\rho=1$ for $d=0.125$ ($c=0.25$, blue) and $d=0.05$ ($c=0.10$, orange); solid branches are stable and dashed branches unstable. The lower fold moves from $b_{a,-}\simeq0.79$ to $0.65$, whereas the upper fold changes only from $b_{a,+}\simeq0.96$ to $0.95$. Parameters: $b_h=1$, $m=0.002$, $\omega=2$, and $\alpha=1$; $\eta=0$ in A and C.}
    \label{fig:diversity}
\end{figure*}

\medskip
\section{Crowding from diversity of AI outputs}. 
The crowding term in \eqref{eq:payoffA} captures the loss of value as AI-assisted outputs become more similar to those produced by others. This mechanism is motivated by evidence that generative AI can reduce the diversity of creative outputs 
\cite{doshi_generative_2024,meincke_chatgpt_2025,moon_homogenizing_2025,ashkinaze_how_2025}. When originality is rewarded, this similarity imposes a frequency-dependent cost: an AI-assisted output becomes less distinctive as similar outputs become more common. 
Here, we show that the crowding coefficient $c$ in \eqref{eq:payoffA}  admits a microscopic 
interpretation in terms of this overlap between outputs, 
and examine how changes in the diversity of AI outputs affect collective dynamics.

Suppose outputs can be assigned to $M$ different  categories. Autonomous and AI-assisted outputs are drawn from distributions $\mathbf p=(p_1,\ldots,p_M)$ and $\mathbf q=(q_1,\ldots,q_M)$, respectively. Two outputs overlap when they fall in the same category, and we assume that such an overlap carries a cost $\omega$ when distinctiveness is rewarded.

Averaging this cost over the strategy of a randomly chosen competitor gives a payoff difference that is linear in the fraction $x$ of AI users. The $x$-dependent part has exactly the form of the crowding term in \eqref{eq:payoffA}, with
\begin{equation}
    c=\omega\|\mathbf q-\mathbf p\|^2 .
    \label{eq:c-general}
\end{equation}
The remaining $x$-independent contribution can be absorbed into the baseline benefit $b_a$ 
of \eqref{eq:payoffA} 
(see SI). Thus, the crowding strength is not an independent abstract cost: it can be directly related to the difference between the distributions of autonomous and AI-assisted outputs.

This relation becomes particularly transparent when autonomous (human) outputs provide a uniform benchmark, $p_j=1/M$. In this case
\begin{equation}
    c=\omega d,
    \qquad
    d=\sum_{j=1}^{M}q_j^2-\frac{1}{M},
    \label{eq:c-from-d}
\end{equation}
where $\sum_j q_j^2$ is the Simpson concentration index of AI-assisted outputs \cite{simpson_measurement_1949}. Hence $d$ measures their excess concentration relative to the autonomous benchmark: $d$ is large when AI outputs are concentrated on relatively few categories and vanishes when they are distributed as broadly as the benchmark population.

Notice that the linear dependence on $x$ we have derived,  corresponds to the simplest case in which overlap with a single competitor matters. 
More generally, the cost of similarity may saturate once similar outputs are already common, or may become significant only when repetition exceeds a threshold. 
These effects can be captured through the exponent $\alpha$ in \eqref{eq:payoffA}: $\alpha<1$ describes a crowding cost that grows at a diminishing rate, whereas $\alpha>1$ describes a cost that remains small at low adoption but rises more sharply as adoption increases.
A microscopic mechanism generating nonlinear crowding is described in the SI.

The relation we have derived, $c=\omega d$, allows to investigate how output diversity affects the feedback identified in the previous sections. In the uniform-human benchmark, making AI outputs more diverse corresponds to reducing $d$, and therefore reduces the crowding force opposing AI adoption. The effective feedback of \eqref{eq:gamma-eff}
reads:
\begin{equation}
\Gamma_{\mathrm{eff}}
    =
    \frac{b_h(1-\eta)}{1+\rho}-\omega d .
    \label{eq:gamma-eff-diversity}
\end{equation}
Thus, reducing the concentration of AI-generated outputs has two opposing effects: it increases the distinctiveness of AI-assisted work, but it also weakens the crowding force that acts as a brake on AI adoption. As a result, the skill-mediated feedback becomes relatively stronger.
Bistability and hysteresis are possible when $\Gamma_{\mathrm{eff}}>8m$, or equivalently
\begin{equation}
    d<
    d_c=
    \frac{1}{\omega}
    \left[
        \frac{b_h(1-\eta)}{1+\rho}-8m
    \right].
    \label{eq:d-critical}
\end{equation}
Figure~\ref{fig:diversity}A shows this boundary. With the parameters shown, strategy--skill correlations reduce the critical value from $d_c\simeq0.49$ for $\rho=0$ to $d_c\simeq0.24$ for $\rho=1$.
The effect on the hysteresis loop is strongly asymmetric. Reducing $d$ has little effect on the upper fold, where the population leaves the low-adoption state, but moves the lower fold substantially toward smaller $b_a$ [Fig.~\ref{fig:diversity}A,C; SI].
Greater output diversity therefore does not strongly promote the initial transition to widespread AI adoption. Instead, once the high-adoption state has been reached, it allows that state to persist at substantially lower AI benefit, making the transition harder to reverse.

The same mechanism determines how much skill must be retained under AI use to eliminate bistability. Increasing $\eta$ weakens the skill-mediated feedback because AI-assisted work preserves a larger fraction of autonomous skill. Above a critical value $\eta_{\mathrm{crit}}$, the two folds merge and the bistable interval disappears, leaving a single equilibrium for every $b_a$ [Fig.~\ref{fig:diversity}B]. As $d$ decreases, crowding becomes weaker and a larger value of $\eta$ is therefore required to eliminate bistability. Output diversity and skill preservation thus act on opposite sides of the same feedback: reducing AI output concentration weakens the crowding force opposing AI adoption, whereas preserving skill weakens the reinforcing effect of skill loss.

This result does not imply that less diverse AI outputs are desirable. Rather, it isolates a single feedback mechanism while holding other quantities fixed. In practice, interventions that alter output diversity may also affect the intrinsic value of AI-assisted work, $b_a$, or the value assigned to originality, $\omega$. The prediction is that evaluating an intervention solely by the diversity of its outputs may overlook its longer-term effects on adoption dynamics.

\section{Conclusions}

We have studied how AI adoption and human skills coevolve when individuals choose between autonomous and AI-assisted work. Skills evolve through practice and disuse, while widespread AI adoption can reduce the distinctiveness of AI-assisted outputs.
Four main results emerge. First, skill erosion can make AI adoption self-reinforcing and generate alternative stable states and hysteresis when this feedback outweighs the opposing effect of crowding. Second, once skill is attached to individuals rather than represented only by a population average, autonomous workers become systematically more skilled than AI users. This strategy--skill correlation partly insulates autonomous work from population-wide skill decline and therefore weakens the feedback driving bistability. Third, the same timescale ratio $\rho=\varepsilon/\lambda$ that controls this correction also determines the distribution of individual skills: frequent strategy changes produce a relatively homogeneous population, whereas persistent strategies can produce a population polarized between highly skilled and strongly deskilled individuals. Finally, relating crowding to output concentration shows that, within the benchmark considered here, reducing the concentration of AI-generated outputs weakens the competitive force restraining adoption and can make a high-adoption state harder to reverse.

These results connect mechanisms that have previously been studied largely in isolation. Tipping points between competence and reliance have been derived for a single user \cite{krakauer_competitive_2026}, while models of cognitive offloading have shown how differences in skill can emerge across independently optimizing workers \cite{caosun_augmentation_2026}. At the population level, social diffusion of AI use coupled to changes in autonomous cognition can itself generate bistability, hysteresis, and persistent dependence \cite{sole_large-language_2026}. Our results identify an additional effect that arises when the coevolving state is individual rather than shared: because each agent's skill records their own history of strategy choices, strategy and skill become correlated. This correlation changes the collective feedback and, at the same time, generates a distribution of skills that cannot be reconstructed from the population mean. The model therefore separates two consequences of AI adoption that may otherwise appear to be the same phenomenon: collective lock-in to widespread adoption and polarization of autonomous skills across individuals. The crowding mechanism, in turn, is motivated by empirical evidence that generative AI can make independently produced outputs more similar \cite{doshi_generative_2024,meincke_chatgpt_2025,moon_homogenizing_2025,azad_ex_2026}.

The timescale ratio $\rho$ provides a direct bridge between the theory and empirical settings. Since
\begin{equation}
    \rho=\frac{\varepsilon}{\lambda}
    =\ln 2\,\frac{\tau}{t_{1/2}},
\end{equation}
where $t_{1/2}=\ln2/\varepsilon$ is the half-life of an unpractised skill and $\tau=1/\lambda$ is the correlation time of strategy choice, $\rho$ can in principle be estimated from two independently measurable quantities. Estimates of skill retention provide information about $t_{1/2}$ \cite{arthur_jr_factors_1998,krakauer_competitive_2026}, while longitudinal records of AI use can reveal how persistently individuals retain the same strategy \cite{contractor_generative_2026}. The theory then predicts specific signatures across settings with different values of this ratio. As $\rho$ increases, the skill gap between current autonomous and AI-assisted workers should increase as $\rho/(1+\rho)$, while the distribution of unaided skill should broaden and eventually become bimodal. Conversely, when AI use is revised frequently, strategy--skill correlations should remain weak and the population-level dynamics should be closer to that predicted from average skill alone. These signatures provide a direct way to test the theory using data that combine longitudinal records of AI use with assessments of unaided performance.
A test that does not require estimating $\rho$ follows from the exact stationary relation $\sigma^\star=-\mathrm{Var}(s)/(1-\eta)$, which holds whatever the switching rule. It implies that the skill gap between current autonomous and AI-assisted workers equals $\mathrm{Var}(s)/[(1-\eta)x^\star(1-x^\star)]$. 
For a population approximately at stationarity and obeying the skill dynamics of \eqref{eq:si-evolution}, cross-sectional measurements of AI use and unaided skill on a common scale can therefore test this relation or, conversely, estimate the skill range $1-\eta$ lost under sustained AI use.
The relative benefit of AI-assisted work, $b_a$, can likewise be informed by experiments, and it is not a fixed property of the technology: blind comparisons of human and AI-generated responses find that AI scores higher on closed, structured tasks but lower on open-ended ones~\cite{charness_creativity_2026}. Different tasks may therefore sit in different regions of the phase diagram of Fig.~\ref{fig:meanfield}C, with open-ended tasks closer to the low-adoption state and closed tasks closer to, or beyond, the bistable region.

The model also distinguishes interventions that act on different parts of the feedback. Preserving practice during AI-assisted work increases the lowest possible skill level $\eta$ and directly weakens the reinforcing link between AI adoption and skill loss; sufficiently high retention eliminates bistability. This is consistent with evidence that forms of AI assistance designed to preserve active engagement can support learning more effectively than simply providing finished answers \cite{bastani_generative_2025}. Rewarding distinctiveness instead strengthens the crowding force opposing widespread adoption. Changes in output diversity act differently again. Within the uniform autonomous-output benchmark, 
making AI-assisted outputs more diverse reduces overlap at a given level of adoption but also weakens the crowding force that restrains further adoption, allowing the high-adoption state to persist at lower intrinsic AI benefit. Improving output diversity and preserving autonomous skill are therefore not equivalent objectives. More generally, the model reveals a second tradeoff controlled by $\rho$: frequent strategy changes limit differences in skill across individuals but leave the collective feedback relatively strong, whereas persistent strategies weaken collective lock-in while increasing skill polarization. Preventing collective lock-in and preventing the concentration of autonomous skill in only part of the population can therefore require different interventions. Because the system can exhibit hysteresis, timing also matters: after the population crosses a fold into the high-adoption state, restoring the parameter that changed need not be sufficient to restore the previous state.

The model is deliberately minimal, and its predictions should therefore be interpreted as consequences of the mechanisms included rather than as universal forecasts of AI use. AI performance and output distributions are held fixed, whereas real systems
continue to evolve and may themselves be trained on increasingly AI-generated material \cite{shumailov_ai_2024,gerstgrasser_is_2024,baumann_generative_2026}.
A natural extension would be to make the distribution of AI outputs itself dynamic through a reinforced urn process, in which both human- and AI-generated outputs contribute to the pool from which future outputs are generated \cite{tria_dynamics_2014,loreto_dynamics_2016,
iacopini_network_2018,di_bona_dynamics_2025}. In the present framework, this would turn the excess-concentration parameter $d$ from a fixed quantity into a dynamical variable coupled to AI adoption, allowing adoption to modify not only human skill but also the future distribution of AI-generated outputs. Skill recovery is always possible under autonomous practice, while in some domains relearning may itself become harder after prolonged disuse \cite{park_enrichment_2026}. 
Strategy revision is based on the observed returns of others, which may reflect the role model's individual skill as well as the strategy itself \cite{livan_dont_2019}; alternative update rules based on counterfactual returns could therefore modify the dynamics. The effective feedback $\Gamma_{\mathrm{eff}}$ is derived in the weak-selection regime and validated here by agent-based simulations, while interactions are assumed to be well mixed. Network structure, heterogeneous tasks, and persistent differences among individuals are natural extensions. Finally, AI-assisted work need not always erode autonomous skill: forms of assistance that actively support learning \cite{lira_coach_2026} could reduce or even reverse the strategy--skill correlation. These limitations also indicate what should be measured to distinguish among regimes in real populations: not only rates of AI adoption and average performance, but the persistence of AI-use choices, the retention of unaided skill, and the full distribution of individual skills. The long-run consequences of AI adoption therefore cannot generally be inferred from present-day performance or population averages alone: they depend on how individual histories of use reshape both the stability of collective adoption and the distribution of human skill.

\section{Methods}
\subsection*{Agent-based model}
\label{methods:agents}
We simulate a well-mixed population of $N$ agents in continuous time. Revision events occur at total rate $N$, so that each agent revises its strategy at unit rate and one sweep corresponds to one revision per agent on average. At a revision, the focal agent switches strategy with probability $\mu$; otherwise it samples a role model uniformly at random and, if the two use different strategies, adopts the role model's strategy with the probability of \eqref{eq:fermi}. The payoff of an autonomous agent is $b_hs_i$. The payoff of an AI user is $b_a-c\,\hat{x}^{\alpha}$, where $\hat{x}=n/k$ and $n$ is the number of AI users among $k$ agents sampled uniformly at random, drawn anew at each payoff evaluation. In a well-mixed population $\hat{x}$ is an unbiased estimate of $x$, so the mean-field payoff is recovered for $k\to\infty$, while finite $k$ introduces sampling noise into payoff comparisons \cite{traulsen_pairwise_2007}.  Between events, skills evolve exactly according to \eqref{eq:si-evolution}: an agent with target $\theta_i=1$ (autonomous) or $\theta_i=\eta$ (AI-assisted) has $s_i(t)=\theta_i+[s_i(t_0)-\theta_i]\,e^{-\varepsilon_{\mathrm{sw}}(t-t_0)}$, updated whenever the agent is involved in an event or a measurement is taken.

\subsection*{Units}
Parameters are reported in the rescaled time used in the mean-field theory.
With $w=\beta/2$, the per-revision exploration probability is $\mu=mw/(1+mw)$, and the skill relaxation rate per sweep is $\varepsilon_{\mathrm{sw}}=(1-\mu)w\,\varepsilon$ (SI Appendix). The timescale ratio $\rho$ is dimensionless and identical in both units.

\subsection*{Quasi-static continuation}
For Fig.~\ref{fig:agent-hysteresis}, $b_a$ takes $61$ equally spaced values between $0.15$ and $1.10$. At each value the population relaxes for $1200$ sweeps, and observables are recorded every $5$ sweeps during $300$ further sweeps. The upward sweep starts from $x=0.02$ with $s_i=1$ for all agents; the downward sweep starts from the final state of the upward one. The width of the hysteresis loop is the difference between the values of $b_a$ at which the two branches cross $x=1/2$, obtained by linear interpolation. Results are averaged
over $3$ independent repetitions.

\subsection*{Measured quantities}
The mean skills $s_h$ and $s_a$ are averaged over the recorded snapshots; states in which either strategy is used by fewer than $10$ agents are excluded from
Fig.~\ref{fig:agent-hysteresis}C. The per-capita switching rates are measured as $r_{+}=n_{H\to A}/T_H$ and $r_{-}=n_{A\to H}/T_A$, where $n_{H\to A}$ and $n_{A\to H}$ count strategy changes during the measurement window and $T_H$ and $T_A$ are the total times spent by agents in each strategy; the measured timescale ratio is $\rho=\varepsilon_{\mathrm{sw}}/(r_{+}+r_{-})$.

\subsection*{Skill distributions}
For Fig.~\ref{fig:agent_capacity_distributions}, runs start from $x=0.4$ with randomly assigned strategies and $s_i=1$. After a burn-in of $1668$ sweeps (panel A) or $600$ sweeps (panels B, C), distributions are accumulated over $1800$ sweeps from snapshots taken every $5$ sweeps, in $60$ bins. The value of $b_a$ in each panel is chosen so that the theory predicts $x^{\star}=0.40$.

\subsection*{Theoretical predictions}
Stationary states of the MFT with covariance closure solve $x(1-x)[b_a-b_h+\Gamma_{\mathrm{eff}}x]+m(1-2x)=0$, with $\Gamma_{\mathrm{eff}}$ from \eqref{eq:gamma-eff} and $\rho=\varepsilon/\lambda^{\star}(x)$ from \eqref{eq:rho}. Folds are the local extrema of the resulting equilibrium curve $b_a(x)$, located on a fine grid and refined by bounded minimization; the closing point $\varepsilon_c$ is the value at which the two extrema merge. The predicted Beta law uses the same $x^{\star}$ and $\rho$ [\eqref{eq:beta-law}]. Derivations are given in the SI Appendix.

\section*{Data Availability}

Code reproducing all simulations and figures is available at
\url{https://github.com/GiacomoFrigerio/AI-Adoption-vs-Human-Skills}.

\bibliography{zotero_biblio}

\clearpage
\onecolumngrid
\begingroup

\let\eqref\originaleqref

\setcounter{section}{0}
\setcounter{subsection}{0}
\setcounter{equation}{0}
\setcounter{figure}{0}
\setcounter{table}{0}

\renewcommand{\thesection}{S\arabic{section}}
\renewcommand{\thesubsection}{\thesection.\arabic{subsection}}
\renewcommand{\thesubsubsection}{\thesubsection.\arabic{subsubsection}}

\makeatletter
\renewcommand{\p@subsection}{}
\renewcommand{\p@subsubsection}{}
\makeatother

\numberwithin{equation}{section}
\renewcommand{\theequation}{S\arabic{section}.\arabic{equation}}
\renewcommand{\thefigure}{S\arabic{figure}}
\renewcommand{\thetable}{S\arabic{table}}

\providecommand{\theHsection}{}
\renewcommand{\theHsection}{supp.\arabic{section}}
\providecommand{\theHequation}{}
\renewcommand{\theHequation}{supp.\arabic{section}.\arabic{equation}}
\providecommand{\theHfigure}{}
\renewcommand{\theHfigure}{supp.\arabic{figure}}
\providecommand{\theHtable}{}
\renewcommand{\theHtable}{supp.\arabic{table}}

\begin{center}
{\Large\bfseries Supplemental Material\par}
\medskip
{\large Modeling the Coevolution of Human Skills and AI Adoption\par}
\end{center}
\medskip

\section{Notation}
\label{supp:sec:notation}

\begin{table}[h]
\centering
\caption{Symbols used throughout. Rates are expressed in the rescaled time of
the mean-field equations unless the subscript ``sw'' (per sweep of the
simulations) is used.}
\label{supp:tab:notation}
\small
\begin{tabular}{@{}ll@{}}
\toprule
Symbol & Meaning\\
\midrule
$N$ & number of agents\\
$z_i\in\{H,A\}$ & strategy of agent $i$: autonomous ($H$) or AI-assisted ($A$)\\
$s_i\in[\eta,1]$ & autonomous skill of agent $i$\\
$u_i=(s_i-\eta)/(1-\eta)\in[0,1]$ & rescaled skill\\
$x$ & fraction of AI users\\
$s$ & population mean skill, $s=N^{-1}\sum_is_i$\\
$s_h,\ s_a$ & mean skill of autonomous agents and of AI users\\
$\sigma$ & strategy--skill covariance, $\sigma=\E[\1(z_i=A)s_i]-xs$\\
$b_h$ & benefit per unit skill of autonomous work, $\pi_H=b_hs_i$\\
$b_a$ & baseline benefit of AI-assisted work, $\pi_A=b_a-cx^{\alpha}$\\
$c,\ \alpha$ & strength and exponent of crowding\\
$\eta$ & skill retained under sustained AI use\\
$\eps$ & rate of skill adaptation\\
$\beta$ & intensity of selection in the Fermi rule\\
$\mu$ & probability of spontaneous switching per revision\\
$m$ & rescaled spontaneous switching rate, $m=2\mu/[(1-\mu)\beta]$\\
$k$ & number of agents sampled to estimate crowding\\
$r_{+},\ r_{-}$ & per-capita switching rates $H\to A$ and $A\to H$\\
$\lambda=r_{+}+r_{-}$ & total switching rate\\
$\rho=\eps/\lambda$ & ratio of skill-adaptation and strategy-switching rates\\
$\Gamma=b_h(1-\eta)-c$ & net feedback coefficient of the reduced MFT\\
$\geff=b_h(1-\eta)/(1+\rho)-c$ & effective feedback coefficient\\
\bottomrule
\end{tabular}
\end{table}

Expectations $\E[\cdot]$ are averages over agents. Conditional averages over
autonomous agents or AI users are written $\E[\cdot\mid H]$,
$\E[\cdot\mid A]$, or $\langle\cdot\rangle_H$, $\langle\cdot\rangle_A$.
Stationary values carry a star, $x^{\star}$.

\section{Microscopic model and simulation algorithm}
\label{supp:sec:abm}

\subsection{Agents, payoffs and skills}

Each agent $i=1,\dots,N$ has a strategy $z_i(t)\in\{H,A\}$ and a skill
$s_i(t)\in[\eta,1]$. Payoffs are defined as
\begin{equation}
    \pi_H(i)=b_hs_i,\qquad \pi_A=b_a-c\,\hat{x}^{\alpha},
    \label{supp:eq:payoffs}
\end{equation}
where $\hat{x}$ is the fraction of AI users estimated from a sample of the
population (Section~\ref{supp:sec:crowding-sample}). 

In this section, $t$ denotes simulation time measured in sweeps. Skills evolve according to
\begin{equation}
    \frac{ds_i}{dt}
    = \varepsilon_{\mathrm{sw}}(\theta_i-s_i),
    \qquad
    \theta_i =
    \begin{cases}
        1, & z_i=H,\\
        \eta, & z_i=A.
    \end{cases}
\end{equation}
Here $\varepsilon_{\mathrm{sw}}$ is the skill-adaptation rate per sweep. The corresponding rate in rescaled time is $\varepsilon=\varepsilon_{\mathrm{sw}}/[(1-\mu)\beta/2]$; the time conversion is derived in Section~\ref{supp:sec:rescaling-time}.

\paragraph{Skill as a memory of past strategies.}
Using the rescaled skill $u_i=(s_i-\eta)/(1-\eta)$ and rescaled time $\tau=(1-\mu)\beta t/2$, the skill dynamics becomes
$$
    \frac{du_i}{d\tau}=\varepsilon(\chi_i-u_i),
    \qquad \chi_i=\mathbf{1}(z_i=H).
$$
After the initial-condition contribution has decayed, its solution is
\begin{equation}
    u_i(\tau)
    =\varepsilon\int_0^\infty
      e^{-\varepsilon\ell}\chi_i(\tau-\ell)\,d\ell.
\end{equation}
Skill is therefore an exponentially weighted record of recent autonomous work, with memory time $1/\varepsilon$ in rescaled units.

\subsection{Strategy revision}

Revision events occur as a Poisson process of total rate $N$ per unit simulation time (one \emph{sweep}), so that each agent revises at rate one. At a revision:
\begin{enumerate}
\item a focal agent $i$ is drawn uniformly at random;
\item with probability $\mu$ it switches strategy (exploration);
\item otherwise it draws a role model $j$ uniformly at random and, if $z_j\neq z_i$, adopts $z_j$ with probability
\begin{equation}
    \phi(\pi_j-\pi_i)=\frac{1}{1+\exp[-\beta(\pi_j-\pi_i)]}.
    \label{supp:eq:fermi}
\end{equation}
\end{enumerate}
In an $H$--$A$ comparison exactly one of the two agents is an AI user, so $\hat{x}$ is sampled once per comparison.

\subsection{Sampled crowding}
\label{supp:sec:crowding-sample}

The AI payoff uses $\hat{x}=n/k$, where $n$ is the number of AI users among $k$ agents sampled uniformly (with replacement in the large-$N$ limit). Then $n\sim\mathrm{Bin}(k,x)$, so
\begin{equation}
    \E[\hat{x}]=x,\qquad
    \mathrm{Var}[\hat{x}]=\frac{x(1-x)}{k}.
    \label{supp:eq:xhat-moments}
\end{equation}
For $\alpha\neq1$, a second-order expansion of $g(\hat{x})=\hat{x}^{\alpha}$ around $x$ gives
\begin{equation}
    \E[\hat{x}^{\alpha}]
    =x^{\alpha}+\tfrac12\,\alpha(\alpha-1)x^{\alpha-2}\,\mathrm{Var}[\hat{x}]+\dots
    =x^{\alpha}+\frac{\alpha(\alpha-1)}{2k}\,x^{\alpha-1}(1-x)+O(k^{-2}),
    \label{supp:eq:xhat-alpha}
\end{equation}
so that the mean-field payoff $b_a-cx^{\alpha}$ is recovered exactly for $\alpha=1$ and up to $O(1/k)$ otherwise. The variance adds noise to payoff comparisons, which effectively lowers the intensity of selection \cite{traulsen_pairwise_2007}; at weak selection it does not enter the deterministic drift (Section~\ref{supp:sec:replicator}) in the linear crowding scenario.

\subsection{Algorithm}

Simulations are carried out through the following steps.
Throughout this algorithm, time is measured in sweeps.

\begin{enumerate}
    \item Initialise $z_i$ and $s_i$; set $t=0$ and
    $t_{0,i}=0$ for every agent.

    \item Draw a waiting time $\Delta t$ from an exponential
    distribution with rate $N$, and set the next revision time
    to $t_{\mathrm{event}}=t+\Delta t$.
    Process any scheduled measurements before this revision,
    as described below, and then set $t=t_{\mathrm{event}}$.

    \item Draw the focal agent $i$ uniformly at random.
    Update its skill to time $t$ according to
    \begin{equation}
        s_i(t)=\theta_i+
        [s_i(t_{0,i})-\theta_i]
        e^{-\varepsilon_{\mathrm{sw}}(t-t_{0,i})},
    \end{equation}
    and set $t_{0,i}=t$.

    \item With probability $\mu$, flip $z_i$, update its
    target $\theta_i$ accordingly, and return to step~2.

    \item Otherwise, draw a role model $j$ uniformly from
    the other $N-1$ agents. If $z_j=z_i$, return to step~2.
    If their strategies differ, update $s_j$ to time $t$
    using the same relaxation formula and set $t_{0,j}=t$.
    Draw $n\sim\mathrm{Bin}(k,x)$, set
    $\pi_A=b_a-c(n/k)^\alpha$, and compute the payoff
    difference in favour of the role model's strategy.
    Set $z_i\leftarrow z_j$ with the Fermi probability,
    updating $\theta_i$ if the strategy changes.
    Return to step~2.
\end{enumerate}

At each scheduled measurement time, update all skills to that time using the exponential relaxation formula, set their timestamps to the measurement time, and record the observables.

\section{Exact equation for the mean skill}
\label{supp:sec:meanskill}

In this section, dots denote derivatives with respect to rescaled time $\tau=(1-\mu)\beta t/2$, and $\varepsilon=\varepsilon_{\mathrm{sw}}/[(1-\mu)\beta/2]$. \\
The population mean skill is $s=N^{-1}\sum_is_i$. Using $\theta_i=1$ for autonomous agents and $\theta_i=\eta$ for AI users, the target can be written as $\theta_i=1-(1-\eta)\1(z_i=A)$. Averaging the skill dynamics over agents in rescaled time,
\begin{align}
    \dot s
    &=\frac1N\sum_{i=1}^N\eps(\theta_i-s_i)
     =\eps\left[\frac1N\sum_i\theta_i-\frac1N\sum_is_i\right]\nonumber\\
    &=\eps\left[1-(1-\eta)\frac1N\sum_i\1(z_i=A)-s\right]
     =\eps\left[1-(1-\eta)x-s\right].
    \label{supp:eq:mean-skill}
\end{align}
This is \Meq{eq:reduced-s}. No closure is needed: because the skill law is linear in $s_i$ and its target depends only on the current strategy, the mean of the targets is a function of $x$ alone, whatever the distribution of individual skills. Strategy switches do not contribute, because a switch
changes $\theta_i$ but not $s_i$. At stationarity
\begin{equation}
    s^{\star}=1-(1-\eta)x^{\star}.
    \label{supp:eq:mean-skill-star}
\end{equation}

The mean skill can be decomposed by strategy,
\begin{equation}
    s=(1-x)s_h+xs_a,\qquad
    s_h=\E[s_i\mid H],\quad s_a=\E[s_i\mid A],
    \label{supp:eq:decomposition}
\end{equation}
since a fraction $1-x$ of agents is autonomous with mean skill $s_h$ and a
fraction $x$ uses AI with mean skill $s_a$.

\section{Strategy dynamics from pairwise comparison}
\label{supp:sec:replicator}

We derive \Meq{eq:replicator-sh} from the revision process, following the standard route from pairwise comparison to replicator dynamics \cite{traulsen_stochastic_2006,traulsen_pairwise_2007}. The only new element is that the autonomous payoff depends on the individual skill.

\subsection{Transition probabilities}

Let $K=\sum_i\1(z_i=A)=Nx$. A revision changes $K$ by at most one, so $x$ changes by $+1/N$ ($H\to A$), $-1/N$ ($A\to H$), or not at all. Let $T^{+}$ and $T^{-}$ be the probabilities of an increase and a decrease in one revision.

\paragraph{Increase ($H\to A$).} The focal agent must be autonomous, with probability $1-x$. It then switches either by exploration with probability $\mu$, or by drawing an AI user as role model and copying it with probability $1-\mu$. When the focal agent is autonomous, all $K$ AI users are among the other $N-1$ agents, so
\begin{equation}
    P(j\in A\mid i\in H)=\frac{K}{N-1}=\frac{Nx}{N-1}.
\end{equation}
The payoff difference in favour of adopting AI is
$\pi_A-\pi_H(i)=b_a-c\hat{x}^{\alpha}-b_hs_i$. Hence
\begin{equation}
    T^{+}=(1-x)\left[\mu+(1-\mu)\frac{Nx}{N-1}
    \left\langle\phi\!\left(b_a-c\hat{x}^{\alpha}-b_hs_i\right)\right\rangle_{H}\right],
    \label{supp:eq:Tplus}
\end{equation}
where the average is over the skill of the focal agent, drawn from the autonomous subpopulation, and over the crowding sample.

\paragraph{Decrease ($A\to H$).} The focal agent must use AI, with probability $x$. It switches by exploration with probability $\mu$, or copies an autonomous role model; since the focal agent is an AI user, all $N-K$ autonomous agents are among the other $N-1$, so $P(j\in H\mid i\in A)=N(1-x)/(N-1)$. The payoff difference in favour of the role model is $\pi_H(j)-\pi_A=b_hs_j-b_a+c\hat{x}^{\alpha}$, with $s_j$ drawn from the autonomous subpopulation:
\begin{equation}
    T^{-}=x\left[\mu+(1-\mu)\frac{N(1-x)}{N-1}
    \left\langle\phi\!\left(b_hs_j-b_a+c\hat{x}^{\alpha}\right)\right\rangle_{H}\right].
    \label{supp:eq:Tminus}
\end{equation}
The AI user's own skill does not appear, because its payoff does not depend on it. In both directions it is the skill of the autonomous member of the pair that enters.

\subsection{Weak-selection expansion}

The Fermi function in Eq.~\eqref{supp:eq:fermi} can be written as $\phi(y)=\tfrac12+\tfrac12\tanh(\beta y/2)$.
With $\tanh z=z-z^3/3+O(z^5)$,
\begin{equation}
    \phi(y)=\frac12+\frac{\beta}{4}\,y-\frac{\beta^3}{48}\,y^3+O(\beta^5y^5).
    \label{supp:eq:fermi-expansion}
\end{equation}
For weak selection, $\beta|y|\ll1$, we keep the linear term. Since $\phi$ is then linear in $y$, averages act only on the argument. Using $\E[\hat{x}]=x$ for $\alpha=1$ [or Eq.~\eqref{supp:eq:xhat-alpha} otherwise] and $\E[s_i\mid H]=s_h$, and defining the mean payoff advantage of AI in an $H$-$A$ comparison,
\begin{equation}
    D=b_a-cx^{\alpha}-b_hs_h,
    \label{supp:eq:D}
\end{equation}
we obtain
\begin{equation}
    \left\langle\phi(\pi_A-\pi_H)\right\rangle_H=\frac12+\frac{\beta}{4}D,
    \qquad
    \left\langle\phi(\pi_H-\pi_A)\right\rangle_H=\frac12-\frac{\beta}{4}D.
    \label{supp:eq:phi-averages}
\end{equation}
Note that the skill of autonomous agents enters only through its conditional mean $s_h$, and not through the population mean $s$.

\subsection{Drift and large-population limit}

The expected change of $x$ in one revision is $\E[\Delta x]=(T^{+}-T^{-})/N$. Substituting Eqs.~\eqref{supp:eq:Tplus},
\eqref{supp:eq:Tminus} and \eqref{supp:eq:phi-averages},
\begin{align}
    T^{+}-T^{-}
    &=\mu(1-x)-\mu x
    +(1-\mu)\frac{N}{N-1}\,x(1-x)
    \left[\left(\frac12+\frac{\beta}{4}D\right)-\left(\frac12-\frac{\beta}{4}D\right)\right]
    \nonumber\\
    &=\mu(1-2x)+(1-\mu)\frac{N}{N-1}\,\frac{\beta}{2}\,x(1-x)\,D.
    \label{supp:eq:Tdiff}
\end{align}
The terms $\tfrac12$ from payoff-neutral imitation cancel: neutral copying moves agents in both directions at equal rates. The factor $x(1-x)$ appears because payoff-driven switching requires the two agents of a pair to use different strategies.

Revisions occur at rate $N$ per sweep, so the mean
waiting time between revisions is $1/N$. In the limit $N\to\infty$, fluctuations vanish and the drift becomes a derivative,
\begin{equation}
    \frac{dx}{dt}=\frac{\E[\Delta x]}{\Delta t}=T^{+}-T^{-}
    =\frac{(1-\mu)\beta}{2}\,x(1-x)\,D+\mu(1-2x),
    \label{supp:eq:xdot-sweep}
\end{equation}
where we used $N/(N-1)\to1$.

\subsection{Rescaling of time}
\label{supp:sec:rescaling-time}
Define the rescaled time and exploration rate
\begin{equation}
    \tau=\frac{(1-\mu)\beta}{2}\,t,\qquad
    m=\frac{2\mu}{(1-\mu)\beta}.
    \label{supp:eq:rescaling}
\end{equation}
Dividing Eq.~\eqref{supp:eq:xdot-sweep} by $(1-\mu)\beta/2$,
\begin{equation}
    \frac{dx}{d\tau}=x(1-x)\left[b_a-cx^{\alpha}-b_hs_h\right]+m(1-2x),
    \label{supp:eq:replicator}
\end{equation}
which is \Meq{eq:replicator-sh}. All rates of the mean-field theory, including $\eps$ and the switching rates of Section~\ref{supp:sec:rates}, are measured in units of $\tau$. A rate $\kappa_{\sw}$ per sweep corresponds to $\kappa=\kappa_{\sw}/[(1-\mu)\beta/2]$ in rescaled units. Inverting Eq.~\eqref{supp:eq:rescaling} for use in simulations, with $w=\beta/2$,
\begin{equation}
    m(1-\mu)w=\mu
    \;\Longrightarrow\;
    \mu=\frac{mw}{1+mw},
    \qquad
    \eps_{\sw}=(1-\mu)\,w\,\eps .
    \label{supp:eq:units}
\end{equation}
For $m=0.002$, this gives $\mu\simeq5.0\times10^{-4}$ at $\beta=0.5$ and $\mu\simeq1.0\times10^{-3}$ at $\beta=1$. Dimensionless ratios of two rates, such as $\rho=\eps/\lambda$ (see~\Meq{eq:rho}), are the same in both units.

For $m=0$, Eq.~\eqref{supp:eq:replicator} is the two-strategy replicator equation $\dot x=x(1-x)(\pi_A-\langle\pi_H\rangle_H)$, with the autonomous payoff averaged over autonomous agents, $\langle\pi_H\rangle_H=b_hs_h$.

\section{Strategy--skill covariance}
\label{supp:sec:covariance}
\subsection{Definition and conditional means}

Since $\1(z_i=A)s_i$ equals $s_i$ for AI users and zero otherwise,
$\E[\1(z_i=A)s_i]=xs_a$. The covariance between strategy and skill is therefore
\begin{equation}
    \sigma=\E[\1(z_i=A)s_i]-\E[\1(z_i=A)]\,\E[s_i]=xs_a-xs .
    \label{supp:eq:first-sigma}
\end{equation}
Substituting $s=(1-x)s_h+xs_a$ from Eq.~\eqref{supp:eq:decomposition},
\begin{equation}
    \sigma=xs_a-x(1-x)s_h-x^2s_a=x(1-x)s_a-x(1-x)s_h=x(1-x)(s_a-s_h),
    \label{supp:eq:sigma-def}
\end{equation}
which is \Meq{eq:workflow-skill-covariance}. The covariance is negative when
autonomous agents are more skilled than AI users. Solving
Eq.~\eqref{supp:eq:decomposition} and Eq.~\eqref{supp:eq:sigma-def} for the conditional
means, using $s_a-s_h=\sigma/[x(1-x)]$ and $s=s_h+x(s_a-s_h)$, we get
\begin{equation}
    s_h=s-\frac{\sigma}{1-x},\qquad s_a=s+\frac{\sigma}{x}.
    \label{supp:eq:conditional-means}
\end{equation}

\subsection{Strategy equation in terms of the mean skill}

Inserting $s_h$ from Eq.~\eqref{supp:eq:conditional-means} into Eq.~\eqref{supp:eq:replicator},
\begin{align}
    \dot x
    &=x(1-x)\left[b_a-cx^{\alpha}-b_h\left(s-\frac{\sigma}{1-x}\right)\right]+m(1-2x)
    \nonumber\\
    &=x(1-x)\left[b_a-cx^{\alpha}-b_hs\right]+b_hx\sigma+m(1-2x),
    \label{supp:eq:replicator-sigma}
\end{align}
which is \Meq{eq:replicator-covariance}. The reduced mean-field theory (reduced MFT) of the main text corresponds to the closure $\sigma=0$. When $\sigma<0$ the correction $b_hx\sigma$ is negative: comparing AI users with autonomous agents who are more skilled than average lowers the rate at which AI spreads.

\subsection{Dynamics of the conditional means}

Let $r_{+}$ and $r_{-}$ be the per-capita rates of $H\to A$ and $A\to H$ switches. In this section we assume that they may depend on the population state but not on an agent's own skill (skill-blind switching), which holds at leading order in $\beta$ (see Section~\ref{supp:sec:rates}). The composition changes as
\begin{equation}
    \dot x=r_{+}(1-x)-r_{-}x .
    \label{supp:eq:x-rates}
\end{equation}
Differentiating conditional means directly is inconvenient because the membership of the two groups changes. We therefore follow the total skill carried by each group,
\begin{equation}
    U_H=(1-x)s_h,\qquad U_A=xs_a,\qquad U_H+U_A=s .
    \label{supp:eq:masses}
\end{equation}

\paragraph{Autonomous group.} $U_H$ changes through three distinct processes.
($i$) Autonomous agents learn ($\dot s_i=\eps(1-s_i)$); averaged over the group and multiplied by its size, this gives $\eps(1-x)(1-s_h)$. ($ii$) Autonomous agents leave at total rate $r_{+}(1-x)$. Since departures do not depend on skill, departing agents are an unbiased sample of the group and carry mean skill $s_h$. The associated skill loss is then $r_{+}(1-x)s_h$. ($iii$) AI users arrive at total rate $r_{-}x$ carrying their mean skill $s_a$. The associated skill gain is then $r_{-}xs_a$. Hence
\begin{equation}
    \dot U_H=\eps(1-x)(1-s_h)-r_{+}(1-x)s_h+r_{-}xs_a .
    \label{supp:eq:UH}
\end{equation}
By the product rule, $\dot U_H=(1-x)\dot s_h-s_h\dot x$. Equating the two expressions and using Eq.~\eqref{supp:eq:x-rates},
\begin{align}
    (1-x)\dot s_h
    &=\eps(1-x)(1-s_h)-r_{+}(1-x)s_h+r_{-}xs_a+s_h\left[r_{+}(1-x)-r_{-}x\right]
    \nonumber\\
    &=\eps(1-x)(1-s_h)+r_{-}x(s_a-s_h).
\end{align}
The terms proportional to $r_{+}$ cancel: unbiased departures change the size of the group but not its mean. Dividing by $1-x$,
\begin{equation}
    \dot s_h=\eps(1-s_h)+r_{-}\frac{x}{1-x}\,(s_a-s_h).
    \label{supp:eq:sh-dot}
\end{equation}

\paragraph{AI group.} Following the same logic, AI users lose skill ($\dot s_i=\eps(\eta-s_i)$), gain agents at rate $r_{+}(1-x)$ with mean skill $s_h$, and lose agents at rate $r_{-}x$ with mean skill $s_a$, resulting in
\begin{equation}
    \dot U_A=\eps x(\eta-s_a)+r_{+}(1-x)s_h-r_{-}xs_a .
    \label{supp:eq:UA}
\end{equation}
With $\dot U_A=x\dot s_a+s_a\dot x$,
\begin{align}
    x\dot s_a
    &=\eps x(\eta-s_a)+r_{+}(1-x)s_h-r_{-}xs_a-s_a\left[r_{+}(1-x)-r_{-}x\right]
    \nonumber\\
    &=\eps x(\eta-s_a)+r_{+}(1-x)(s_h-s_a),
\end{align}
so that
\begin{equation}
    \dot s_a=\eps(\eta-s_a)+r_{+}\frac{1-x}{x}\,(s_h-s_a).
    \label{supp:eq:sa-dot}
\end{equation}

\subsection{Dynamics of the covariance}

Since $\sigma=U_A-xs$ (see Eq~\eqref{supp:eq:decomposition}),
\begin{equation}
    \dot\sigma=\dot U_A-\dot x\,s-x\dot s .
\end{equation}
We rewrite Eq.~\eqref{supp:eq:UA} using $(1-x)s_h=s-U_A$ and $xs_a=U_A$:
\begin{equation}
    \dot U_A=\eps\eta x-\eps U_A+r_{+}(s-U_A)-r_{-}U_A .
\end{equation}
With Eqs.~\eqref{supp:eq:x-rates} and \eqref{supp:eq:mean-skill},
\begin{align}
    \dot\sigma
    &=\eps\eta x-\eps U_A+r_{+}s-r_{+}U_A-r_{-}U_A
      -s\left[r_{+}(1-x)-r_{-}x\right]
      -\eps x\left[1-(1-\eta)x-s\right].
\end{align}
Let us collect the switching terms and the learning terms separately. the switching terms read
\begin{equation}
    r_{+}s-r_{+}U_A-r_{-}U_A-r_{+}s+r_{+}sx+r_{-}sx
    =-(r_{+}+r_{-})(U_A-xs)=-\lambda\sigma .
\end{equation}
The learning terms read
\begin{align}
    \eps\eta x-\eps U_A-\eps x+\eps(1-\eta)x^2+\eps xs
    &=-\eps(U_A-xs)-\eps(1-\eta)x+\eps(1-\eta)x^2
    \nonumber\\
    &=-\eps\sigma-\eps(1-\eta)x(1-x).
\end{align}
Adding the two, we get
\begin{equation}
    \boxed{\;\dot\sigma=-\eps \bigg[ \left(1+\frac{\lambda}{\eps}\right)\,\sigma+(1-\eta)\,x(1-x) \bigg],
    \qquad \lambda=r_{+}+r_{-}.\;}
    \label{supp:eq:sigma-dot}
\end{equation}
This is \Meq{eq:C_dynamics}. The first term relaxes $\sigma$ to zero at rate $\eps+\lambda$: switching mixes the two groups and skill relaxation erases old differences. The second term is a source, which is negative since $0<x<1$ and $0 \le \eta<1$: learning under autonomous work and losing skill under AI use push the two groups apart at rate $\eps(1-\eta)$.

\subsection{Stationary covariance and skill gap}

Setting $\dot\sigma=0$ in Eq.~\eqref{supp:eq:sigma-dot}, we get
\begin{equation}
    \sigma^{\star}=-\frac{(1-\eta)\,x^{\star}(1-x^{\star})}{1+\lambda^{\star}/\eps},
    \qquad
    s_h^{\star}-s_a^{\star}=-\frac{\sigma^{\star}}{x^{\star}(1-x^{\star})}
    =\frac{1-\eta}{1+\lambda^{\star}/\eps}
    =(1-\eta)\frac{\rho}{1+\rho},
    \label{supp:eq:gap}
\end{equation}
with $\rho=\eps/\lambda^{\star}$. This is \Meq{eq:C_stationary}.

The same result follows from the conditional means. At stationarity, Eqs.~\eqref{supp:eq:sh-dot} and \eqref{supp:eq:sa-dot} give
\begin{equation}
    0=\eps(1-s_h)-r_{-}\frac{x}{1-x}(s_h-s_a),\qquad
    0=\eps(\eta-s_a)+r_{+}\frac{1-x}{x}(s_h-s_a).
\end{equation}
The balance of flows $r_{+}(1-x)=r_{-}x$ gives $r_{-}x/(1-x)=r_{+}$ and $r_{+}(1-x)/x=r_{-}$, so
\begin{equation}
    \eps(1-s_h)=r_{+}(s_h-s_a),\qquad \eps(s_a-\eta)=r_{-}(s_h-s_a).
\end{equation}
Adding the two equations, $\eps\left[(1-\eta)-(s_h-s_a)\right]=(r_{+}+r_{-})(s_h-s_a)$, gives Eq.~\eqref{supp:eq:gap} again.

\paragraph{Limits.} For $\rho\to0$, $s_h^{\star}-s_a^{\star}\to0$: agents switch many times before their skill changes, and the two groups are statistically identical. For $\rho\to\infty$, $s_h^{\star}\to1$ and $s_a^{\star}\to\eta$: each group relaxes fully to its own target.

\paragraph{Transients and the sign of $\sigma$.} 
Eq.~\eqref{supp:eq:sigma-dot} is linear in $\sigma$. For fixed $x$ and $\lambda$ its solution is
$\sigma(t)=\sigma^{\star}+[\sigma(0)-\sigma^{\star}]e^{-(\eps + \lambda)t}$. A perturbation that makes AI users more skilled than autonomous agents, for example the adoption of AI by the most skilled agents, produces a transiently positive $\sigma$, which relaxes to the negative stationary value on the timescale $1/(\lambda + \eps)$. The sign of $\sigma^{\star}$ is fixed by the source term and hence by $\eta\le1$. If AI-assisted work sustained more skill than autonomous practice, the source would change sign and AI users would be, at stationarity, the more skilled group.

\section{Switching rates and the timescale ratio}
\label{supp:sec:rates}
\subsection{Per-capita rates at weak selection}
During each sweep, an autonomous agent with skill $s$ switches to AI if they explore (probability $\mu$) or, with probability $1-\mu$, meet an AI user (probability $x$) and copy their strategy. Using Eq.~\eqref{supp:eq:fermi-expansion},
\begin{align}
    r_{+,\sw}(s)
    &=\mu+(1-\mu)\,x\left[\frac12+\frac{\beta}{4}\left(b_a-cx^{\alpha}-b_hs\right)\right]
    \nonumber\\
    &=\mu+(1-\mu)\,x\left[\frac12+\frac{\beta}{4}\left(D-b_h(s-s_h)\right)\right],
    \label{supp:eq:rplus}
\end{align}
where we added and subtracted $b_hs_h$. Similarly, an AI user switches back if they explore or meet an autonomous agent (probability $1-x$) and copy them:
\begin{equation}
    r_{-,\sw}=\mu+(1-\mu)(1-x)\left[\frac12-\frac{\beta}{4}D\right].
    \label{supp:eq:rminus}
\end{equation}
Averaging Eq.~\eqref{supp:eq:rplus} over autonomous agents removes the term $s-s_h$:
\begin{equation}
    \langle r_{+,\sw}\rangle_H=\mu+(1-\mu)\,x\left[\frac12+\frac{\beta}{4}D\right].
    \label{supp:eq:rplus-mean}
\end{equation}
Eq.~\eqref{supp:eq:x-rates} with these rates reproduces Eq.~\eqref{supp:eq:xdot-sweep}:
\begin{align}
    \langle r_{+,\sw}\rangle_H(1-x)-r_{-,\sw}x
    &=\mu(1-x)-\mu x+(1-\mu)x(1-x)\left[\frac{\beta}{4}D+\frac{\beta}{4}D\right]
    \nonumber\\
    &=\mu(1-2x)+\frac{(1-\mu)\beta}{2}\,x(1-x)D .
\end{align}

\paragraph{Skill dependence.} The own skill of an autonomous agent enters
Eq.~\eqref{supp:eq:rplus} only through
\begin{equation}
    r_{+,\sw}(s)-\langle r_{+,\sw}\rangle_H=-(1-\mu)\,x\,\frac{\beta}{4}\,b_h\,(s-s_h),
    \label{supp:eq:rplus-skill}
\end{equation}
which is of order $\beta$. At leading order, switching is skill-blind, as assumed in Section~\ref{supp:sec:covariance}. More skilled autonomous agents adopt AI less often; the consequences are derived in Section~\ref{supp:sec:skill-dependent}.

\subsection{Total switching rate}

Adding Eqs.~\eqref{supp:eq:rplus-mean} and \eqref{supp:eq:rminus}, we get
\begin{align}
    \lambda_{\sw}
    &=2\mu+(1-\mu)\left[\frac{x}{2}+\frac{1-x}{2}\right]
      +(1-\mu)\frac{\beta}{4}D\left[x-(1-x)\right]
    \nonumber\\
    &=\frac{1-\mu}{2}+2\mu+\frac{(1-\mu)\beta}{4}\,(2x-1)\,D .
    \label{supp:eq:lambda-D}
\end{align}
At a stationary state, Eq.~\eqref{supp:eq:replicator} gives
$x(1-x)D^{\star}=-m(1-2x^{\star})$, i.e. $D^{\star}=-m(1-2x^{\star})/[x^{\star}(1-x^{\star})]$. Substituting, and using $(1-\mu)\beta m/4=\mu/2$ from Eq.~\eqref{supp:eq:rescaling}, we obtain
\begin{equation}
    \lambda^{\star}_{\sw}
    =\frac{1-\mu}{2}+2\mu+\frac{(1-\mu)\beta}{4}\,m\,\frac{(1-2x^{\star})^2}{x^{\star}(1-x^{\star})}
    =\frac{1-\mu}{2}+2\mu+\frac{\mu}{2}\,\frac{(1-2x^{\star})^2}{x^{\star}(1-x^{\star})}.
    \label{supp:eq:lambda-sweep}
\end{equation}
Converting to rescaled time by dividing by $(1-\mu)\beta/2$, and using $2\mu/[(1-\mu)\beta/2]=2m$ and $(\mu/2)/[(1-\mu)\beta/2]=m/2$,
\begin{equation}
    \boxed{\;\lambda^{\star}=\frac1\beta+2m+\frac m2\,
    \frac{(1-2x^{\star})^2}{x^{\star}(1-x^{\star})}\simeq\frac1\beta+2m.\;}
    \label{supp:eq:lambda-star}
\end{equation}
The total rate is set by payoff-neutral imitation (the term $1/\beta$, i.e.\ $1/2$ per sweep) and by exploration. Payoffs enter only through the last term, which is small for interior equilibria. Near the boundaries it is not: on the low-adoption branch, for instance, $x^{\star}\simeq m/(b_h-b_a)$ (Section~\ref{supp:sec:asymptotics}), and the last term tends to $(b_h-b_a)/2$, which can be a sizeable fraction of $1/\beta$.

\subsection{Interpretation of the rates}

At stationarity the flows balance, $r_{+}(1-x^{\star})=r_{-}x^{\star}$, so
\begin{equation}
    r_{+}=\lambda x^{\star},\qquad r_{-}=\lambda(1-x^{\star}).
\end{equation}
The two per-capita rates differ unless $x^{\star}=1/2$: the rarer strategy has the larger per-capita outflow. The mean durations of spells of autonomous and AI-assisted work are $1/r_{+}$ and $1/r_{-}$, and $1/\lambda$ is the correlation time of an agent's strategy. For a two-state Markov process with rates $r_{+}$ and $r_{-}$, the autocorrelation of $\1(z_i=A)$ decays as $\mathrm{e}^{-\lambda t}$ (see \cite{werner_horsthemke_noise-induced_1984, bena_dichotomous_2006}).

\subsection{Timescale ratio}

The ratio $\rho=\eps/\lambda^{\star}$ compares skill adaptation with strategy switching. From Eq.~\eqref{supp:eq:lambda-star},
\begin{equation}
    \rho=\frac{\eps}{1/\beta+2m+\dots}\simeq\beta\eps\quad(m\ll1/\beta).
    \label{supp:eq:rho-beta}
\end{equation}
It is therefore not a free parameter: it is fixed by the rate of skill adaptation and the intensity of selection. In simulation units, $\rho=\eps_{\sw}/\lambda^{\star}_{\sw}$ with $\lambda^{\star}_{\sw}\simeq1/2$.

\section{Reduced mean-field theory}
\label{supp:sec:meanfield}

With the closure $\sigma=0$, the dynamics reduces to (see \Meqs{eq:reduced-x}--\Meqs{eq:reduced-s} of the main text)
\begin{align}
    \dot x&=F(x,s)\equiv x(1-x)\left[b_a-cx^{\alpha}-b_hs\right]+m(1-2x),
    \label{supp:eq:F}\\
    \dot s&=G(x,s)\equiv\eps\left[1-(1-\eta)x-s\right].
    \label{supp:eq:G}
\end{align}
We first treat linear crowding, $\alpha=1$; nonlinear crowding is treated in Section~\ref{supp:sec:nonlinear}.

\subsection{Reduction to one dimension and stability}

Every equilibrium lies on the skill nullcline $s = 1 - (1 - \eta) x $. Substituting into $F$ defines the reduced function
\begin{equation}
        \Phi(x)=F\bigl(x,\bar s(x)\bigr) =x(1-x)\left[b_a-cx^{\alpha}-b_h+b_h(1-\eta)x\right]+m(1-2x),
    \label{supp:eq:Phi}
\end{equation}
whose zeros are the equilibrium values of $x$. For $\alpha=1$ the term in square bracket is linear in $x$, resulting in, 
\begin{equation}
        b_a-cx-b_h+b_h(1-\eta)x=b_a-b_h+\Gamma x,
    \qquad \Gamma=b_h(1-\eta)-c,
    \label{supp:eq:Gamma}
\end{equation}
which is \Meq{eq:gamma-feedback}. Let us point out that an interior equilibrium $(x^{\star},\bar s(x^{\star}))$ is a saddle if $\Phi'(x^{\star})>0$ and is linearly stable if $\Phi'(x^{\star})<0$.

This follows from the calculation of the Jacobian of Eqs.~\eqref{supp:eq:F}--\eqref{supp:eq:G}:
\begin{equation}
    J=\begin{pmatrix}F_x & F_s\\ G_x & G_s\end{pmatrix}
    =\begin{pmatrix}F_x & -b_hx(1-x)\\ -\eps(1-\eta) & -\eps\end{pmatrix},
\end{equation}
whose determinant is
\begin{equation}
    \det J=-\eps F_x-\eps(1-\eta)\,b_hx(1-x)=-\eps\left[F_x+b_h(1-\eta)x(1-x)\right].
\end{equation}
By the chain rule, $\Phi'(x)=F_x+F_s\,\bar s'(x)=F_x+[-b_hx(1-x)][-(1-\eta)]=F_x+b_h(1-\eta)x(1-x)$, hence
\begin{equation}
    \det J=-\eps\,\Phi'(x^{\star}),\qquad
    \mathrm{tr}\,J=F_x-\eps=\Phi'(x^{\star})-b_h(1-\eta)x^{\star}(1-x^{\star})-\eps .
\end{equation}
If $\Phi'>0$, then $\det J<0$ and the equilibrium is a saddle. If $\Phi'<0$, then $\det J>0$ and $\mathrm{tr}\,J<0$, since the last two terms of the trace are non-positive; both eigenvalues have negative real part.

Note that the stability analysis is independent of $\eps$, since it affects trajectories and basins of attraction but not the location or stability of equilibria.

\subsection{No spontaneous switching}

For $m=0$ and $\alpha=1$,
\begin{equation}
    \Phi(x)=x(1-x)\left(b_a-b_h+\Gamma x\right).
    \label{supp:eq:Phi_sol}
\end{equation}

\paragraph{Boundary equilibria.} $x=0$ with $s=1$ ($H$) and $x=1$ with $s=\eta$ ($A$) are always equilibria. At $H$, $F_x=(1-2x)[b_a-cx-b_hs]-cx(1-x)$ evaluated at $(0,1)$ gives $F_x=b_a-b_h$, and $F_s=0$; the eigenvalues are $b_a-b_h$ and $-\eps$, so $H$ is stable if and only if
\begin{equation}
    b_a<b_h .
\end{equation}
At $A$, $F_x=-[b_a-c-b_h\eta]$ and $F_s=0$; the eigenvalues are $-(b_a-c-b_h\eta)$ and $-\eps$, so $A$ is stable if and only if
\begin{equation}
    b_a>b_h\eta+c .
\end{equation}

\paragraph{Interior equilibrium.} For $\Gamma\ne0$, the term containing $\Gamma$ in Eq.~\eqref{supp:eq:Phi_sol} vanishes when
\begin{equation}
    x^{\star}=\frac{b_h-b_a}{\Gamma},\qquad s^{\star}=1-(1-\eta)x^{\star},
\end{equation}
which is admissible if $0<x^{\star}<1$. There, we have $\Phi'(x^{\star})=x^{\star}(1-x^{\star})\,\Gamma$. By the stability analysis done in the previous subsection, the interior equilibrium is a saddle for $\Gamma>0$ and stable for $\Gamma<0$.

\paragraph{Regimes.} The three conditions described above are summarised in Table~\ref{supp:tab:regimes}. For $\Gamma>0$, $b_h\eta+c<b_h$, and in the interval $b_h\eta+c<b_a<b_h$ both boundary states are stable and the saddle lies at $0<x^{\star}<1$: the system is bistable, with an interval of width $\Gamma$. For $\Gamma<0$, the interval $b_h<b_a<b_h\eta+c$ has both boundaries unstable and the interior equilibrium stable, resulting in the stable coexistence of strategies.

\begin{table}[h]
\centering
\caption{Stable outcomes of the reduced MFT for $m=0$ and $\alpha=1$.}
\label{supp:tab:regimes}
\small
\begin{tabular}{@{}ll@{}}
\toprule
Region & Stable outcome(s)\\
\midrule
$b_a<\min\{b_h,\,b_h\eta+c\}$ & autonomous state $H$\\
$b_a>\max\{b_h,\,b_h\eta+c\}$ & AI-assisted state $A$\\
$b_h\eta+c<b_a<b_h$ ($\Gamma>0$) & $H$ and $A$, separated by an interior saddle\\
$b_h<b_a<b_h\eta+c$ ($\Gamma<0$) & stable interior coexistence\\
\bottomrule
\end{tabular}
\end{table}

\subsection{Spontaneous switching: the equilibrium curve}

For $m>0$, $\Phi(0)=m>0$ and $\Phi(1)=-m<0$: the flow points into the interval at both boundaries, and all equilibria are interior. For $0<x<1$ we can divide $\Phi(x)=0$ by $x(1-x)$ and solve for $b_a$:
\begin{equation}
    b_a^{\mathrm{eq}}(x)=b_h-\Gamma x+m\,\frac{2x-1}{x(1-x)} .
    \label{supp:eq:ba-curve}
\end{equation}
This curve gives the value of $b_a$ for which a stationary state with adoption $x$ exists. Moreover, $\Phi$ can be written as
\begin{equation}
    \Phi(x)=x(1-x)\left[b_a-b_a^{\mathrm{eq}}(x)\right],
\end{equation}
as one verifies by expanding the right-hand side. At an equilibrium the bracket vanishes, so $\Phi'(x^{\star})=-x^{\star}(1-x^{\star})\,b_a^{\mathrm{eq}\,\prime}(x^{\star})$. Following the very same logic of the stability analysis above, 
branches with $b_a^{\mathrm{eq}\,\prime}(x)>0$ are stable, while branches with $b_a^{\mathrm{eq}\,\prime}(x)<0$ are saddles.

\subsection{Folds and the condition for bistability}

A fold is a turning point of the equilibrium curve, where a stable branch and a saddle branch meet. Differentiating Eq.~\eqref{supp:eq:ba-curve}, and posing $h(x)=(2x-1)/(x-x^2)$, we get
\begin{equation}
    h'(x)=\frac{2(x-x^2)-(2x-1)(1-2x)}{(x-x^2)^2}
    =\frac{2x-2x^2+(1-2x)^2}{x^2(1-x)^2}
    =\frac{1-2x+2x^2}{x^2(1-x)^2}
    =\frac{1}{x^2}+\frac{1}{(1-x)^2},
\end{equation}
where the last step uses $x^2+(1-x)^2=1-2x+2x^2$. Hence
\begin{equation}
    b_a^{\mathrm{eq}\,\prime}(x)=-\Gamma+m\left[\frac1{x^2}+\frac1{(1-x)^2}\right].
    \label{supp:eq:fold-condition}
\end{equation}
The function $q(x)=x^{-2}+(1-x)^{-2}$ diverges at both ends, and has first derivative $q'(x)=-2x^{-3}+2(1-x)^{-3}$, which vanishes only at $x=1/2$, and is convex ($q''(x)>0$). Its minimum is $q(1/2)=8$. Therefore:
\begin{itemize}
\item if $\Gamma<8m$, then $b_a^{\mathrm{eq}\,\prime}(x)>0$ for all $x$: the curve is monotonic, there is a single stable equilibrium for every $b_a$;
\item if $\Gamma>8m$, then $b_a^{\mathrm{eq}\,\prime}(x)<0$ on an interval around $x=1/2$ bounded by two roots $x_{+}<1/2<x_{-}$ of Eq.~\eqref{supp:eq:fold-condition}: the curve has a local maximum $b_{a,+}=b_a^{\mathrm{eq}}(x_{+})$ and a local minimum $b_{a,-}=b_a^{\mathrm{eq}}(x_{-})$, and for $b_{a,-}<b_a<b_{a,+}$ there are two
stable equilibria separated by a saddle.
\end{itemize}
Hence bistability exists for some $b_a$ if and only if
\begin{equation}
    \Gamma>8m,\qquad\text{i.e.}\qquad b_h(1-\eta)-c>8m,
    \label{supp:eq:bistability}
\end{equation}
which is \Meq{eq:bistability-condition}.

\subsection{Cusp}

At $\Gamma=8m$ the two roots merge at $x_c=1/2$. The critical switching rate is $m_c=\Gamma/8$, and the corresponding value of $b_a$ is
\begin{equation}
    b_{a,c}=b_a^{\mathrm{eq}}(1/2)=b_h-\frac{\Gamma}{2}
    =\frac{2b_h-b_h+b_h\eta+c}{2}=\frac{b_h(1+\eta)+c}{2},
\end{equation}
since the switching term vanishes at $x=1/2$. The point $(x_c,m_c,b_{a,c})=(1/2,\Gamma/8,[b_h(1+\eta)+c]/2)$ is a cusp: in the $(b_a,m)$ plane the two fold lines $b_{a,\pm}(m)$ meet there. Equivalently, at fixed $m$ the bistable region in the $(c,b_a)$ plane closes at $c^{\star}=b_h(1-\eta)-8m$ [\Mfig{fig:meanfield}, panel C].

\subsection{Fold positions for small switching}
\label{supp:sec:asymptotics}

For $m\ll\Gamma$, the roots of Eq.~\eqref{supp:eq:fold-condition} lie close to the
boundaries. Near $x=0$, $q(x)\simeq x^{-2}$, so $x_{+}\simeq\sqrt{m/\Gamma}$; near
$x=1$, posing $y=1-x$ results in $y_{-}\simeq\sqrt{m/\Gamma}$.

\paragraph{Upper fold.} For small $x$, $m(2x-1)/[x(1-x)]\simeq-m/x$, so
$b_a^{\mathrm{eq}}(x)\simeq b_h-\Gamma x-m/x$. This is maximal where
$-\Gamma+m/x^2=0$, i.e.\ at $x_{+}=\sqrt{m/\Gamma}$, with value
\begin{equation}
    b_{a,+}\simeq b_h-\Gamma\sqrt{m/\Gamma}-m\sqrt{\Gamma/m}=b_h-2\sqrt{m\Gamma}.
\end{equation}

\paragraph{Lower fold.} For $x=1-y$ with small $y$,
$m(2x-1)/[x(1-x)]=m(1-2y)/[(1-y)y]\simeq m/y$, so
$b_a^{\mathrm{eq}}\simeq b_h-\Gamma+\Gamma y+m/y$, minimal at $y_{-}=\sqrt{m/\Gamma}$:
\begin{equation}
    b_{a,-}\simeq b_h-\Gamma+2\sqrt{m\Gamma}=b_h\eta+c+2\sqrt{m\Gamma}.
\end{equation}
The width of the bistable interval is $b_{a,+}-b_{a,-}\simeq\Gamma-4\sqrt{m\Gamma}$,
which tends to width $\Gamma$ as $m\to0$, with corrections that are $O(\sqrt{m})$.
At $b_h=1$, $c=0.25$, $\eta=0$, $m=0.002$ ($\Gamma=0.75$), these formulas give
$0.923$ and $0.327$, against the exact values $0.925$ and $0.325$.

\paragraph{Stationary adoption near the boundaries.} On the low-adoption branch,
$x^{\star}\ll1$ and $x^{\star}(b_a-b_h)+m\simeq0$, so
$x^{\star}\simeq m/(b_h-b_a)$; on the high-adoption branch,
$1-x^{\star}\simeq m/(b_a-b_h\eta-c)$.

\subsection{Critical crowding and critical retention}

Condition~\eqref{supp:eq:bistability} can be solved for any single parameter. For the crowding strength,
\begin{equation}
    c<c^{\star}=b_h(1-\eta)-8m;
\end{equation}
for the retained skill,
\begin{equation}
    \eta<\eta_{\mathrm{crit}}=\max\left\{0,\;1-\frac{c+8m}{b_h}\right\}.
    \label{supp:eq:eta-crit}
\end{equation}
For $\eta\ge\eta_{\mathrm{crit}}$ no value of $b_a$ supports two stable states.

\subsection{Absence of periodic orbits}

Let $B(x,s)=1/[x(1-x)]$, positive in the interior. Then
\begin{equation}
    \frac{\partial(BF)}{\partial x}+\frac{\partial(BG)}{\partial s}
    =\frac{\partial}{\partial x}\left[b_a-cx^{\alpha}-b_hs+m\frac{1-2x}{x(1-x)}\right]
    -\frac{\eps}{x(1-x)} .
\end{equation}
The first derivative is $-c\alpha x^{\alpha-1}-m\,q(x)$, with $q$ as above, since $\frac{d}{dx}\frac{1-2x}{x(1-x)}=-h'(x)=-q(x)$. Therefore
\begin{equation}
    \frac{\partial(BF)}{\partial x}+\frac{\partial(BG)}{\partial s}
    =-c\alpha x^{\alpha-1}-m\,q(x)-\frac{\eps}{x(1-x)}<0
\end{equation}
throughout the interior. By the Dulac Criterion \cite{steven_h_strogatz_nonlinear_2014}, the reduced MFT has no periodic orbits in the interior of the unit square, for any $\alpha > 0$. All trajectories approach equilibria, and history dependence can only arise from the coexistence of alternative stable equilibria.

\subsection{Hysteresis under a quasi-static protocol}

Suppose $b_a$ is varied slowly compared with the relaxation time of the dynamics. Starting on the low-adoption branch and increasing $b_a$, the state follows this branch, since it remains stable, until the branch ends at the fold $b_{a,+}$; beyond it, the only attractor is the high-adoption branch, and the system jumps there. Decreasing $b_a$ from the high-adoption branch, the state remains on that branch until its own fold $b_{a,-}$. Since $b_{a,-}<b_{a,+}$, the state at a given $b_a$ in $(b_{a,-},b_{a,+})$ depends on the direction of the sweep. The same loop appears in the mean skill through $s^{\star}=1-(1-\eta)x^{\star}$. In finite populations, fluctuations can trigger the jumps before the folds are reached (Section~\ref{supp:sec:finite}).

Because the basin boundary is the stable manifold of the saddle, a curve in the $(x,s)$ plane, the outcome from a given initial state depends on both initial adoption and initial skill [\Mfig{fig:meanfield}, panel A]. A temporary increase of $b_a$ that drives the state across this curve has a permanent effect even if $b_a$ is then restored.

\section{Nonlinear crowding}
\label{supp:sec:nonlinear}

The linear specification $cx$ assumes that each additional AI user adds the same marginal crowding cost. To test this assumption, we retain the skill dynamics and replace the AI payoff by $\pi_A=b_a-cx^\alpha$, with $\alpha>0$. The case $\alpha>1$ represents crowding that becomes stronger at high adoption, whereas $0<\alpha<1$ represents a penalty that rises steeply at low adoption and then saturates. Throughout this section we examine the reduced MFT; the effect of finite sampling on $x^\alpha$ is considered below.

\subsection{Stationary payoff geometry and the new attractors}

Write $B=b_h(1-\eta)$. On the skill nullcline
$s^\star=1-(1-\eta)x^\star$, the stationary payoff advantage of AI is
\begin{equation}
\Delta\pi_\alpha(x)=b_a-b_h+b_h(1-\eta)x-cx^\alpha,
\qquad
\Gamma_\alpha(x)=\frac{d\Delta\pi_\alpha}{dx}
=b_h(1-\eta)-c\alpha x^{\alpha-1}.
\label{supp:eq:nonlinear-payoff-geometry}
\end{equation}
The local coefficient $\Gamma_\alpha(x)$ measures the reinforcement due to skill loss minus the marginal crowding penalty. Unlike the constant $\Gamma=b_h(1-\eta)-c$ of the linear model, it can change sign across the population. Figure~\ref{supp:fig:SI-nonlinear-payoff} displays the two resulting geometries: the payoff difference is concave for $\alpha>1$ and convex for $0<\alpha<1$.

\begin{figure}[t]
\centering
\includegraphics[width=.96\linewidth]{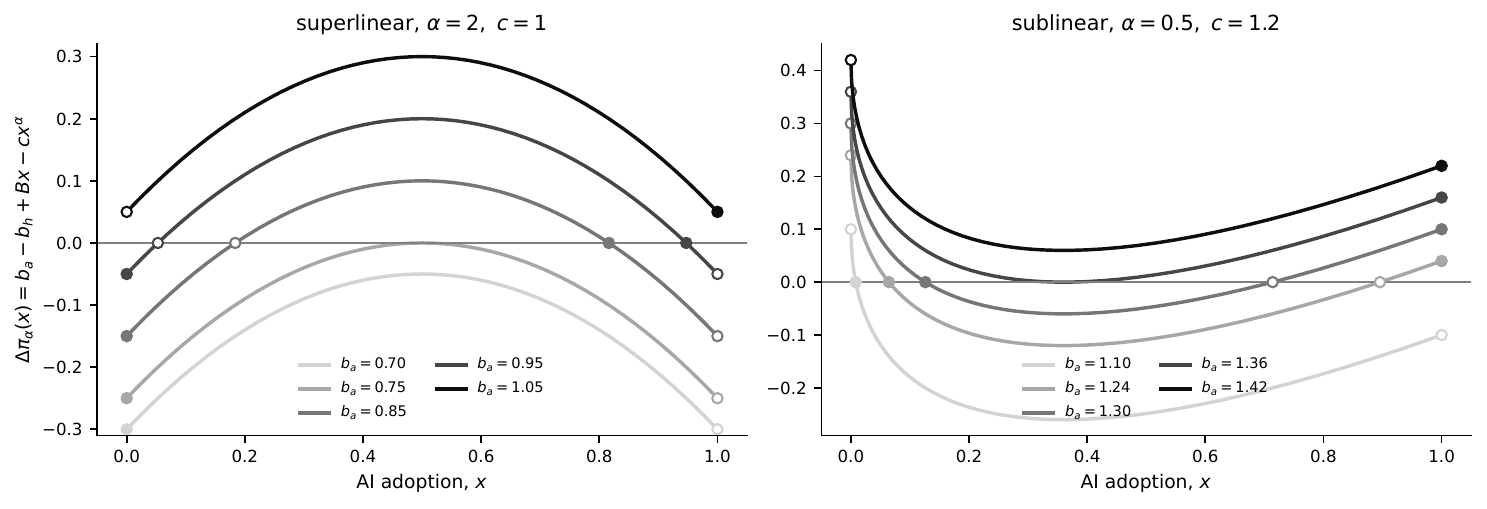}
\caption{\textbf{Stationary payoff difference under nonlinear crowding.}
$\Delta\pi_\alpha(x)$ along the skill nullcline for
$b_h=1$, $\eta=0$ and $m=0$. Left: $\alpha=2$, $c=1$, with
$b_a=0.70,0.75,0.85,0.95,1.05$; the concave payoff difference can have two interior zeros while the autonomous boundary remains stable. Right: $\alpha=1/2$, $c=1.2$, with $b_a=1.10,1.24,1.30,1.36,1.42$; the convex payoff difference can have two interior zeros while the AI boundary remains stable. Filled and open markers denote stable and unstable equilibria, respectively.}
\label{supp:fig:SI-nonlinear-payoff}
\end{figure}

For $m=0$, the boundary states are $H=(0,1)$ and $A=(1,\eta)$. Their stability thresholds, $b_a<b_h$ for $H$ and $b_a>b_h\eta+c$ for $A$, do not depend on $\alpha$. The interior extrema of $\Delta\pi_\alpha$ occur, when $0<x_m<1$, at
\begin{equation}
x_m=\left(\frac{b_h(1-\eta)}{c\alpha}\right)^{1/(\alpha-1)},
\qquad
b_{a,f}=b_h-\left(1-\frac1\alpha\right)b_h(1-\eta)x_m.
\label{supp:eq:nonlinear-interior-fold}
\end{equation}
Here $b_{a,f}$ is the value at which two interior equilibria meet.
For $\alpha>1$ and $c\alpha>B$, an autonomous attractor can coexist with a stable interior state (denoted $H+I$). For $0<\alpha<1$ and $c\alpha<B$, a stable interior state can instead coexist with the near-complete AI state ($I+A$). The attractor map in Fig.~\ref{supp:fig:SI-nonlinear-map} shows both regions, alongside the $H+A$ bistability already present with linear crowding. In particular, the new regions can exist when the average linear coefficient $\Gamma=b_h(1-\eta)-c$ is zero or negative.

\begin{figure}[t]
\centering
\includegraphics[width=.96\linewidth]{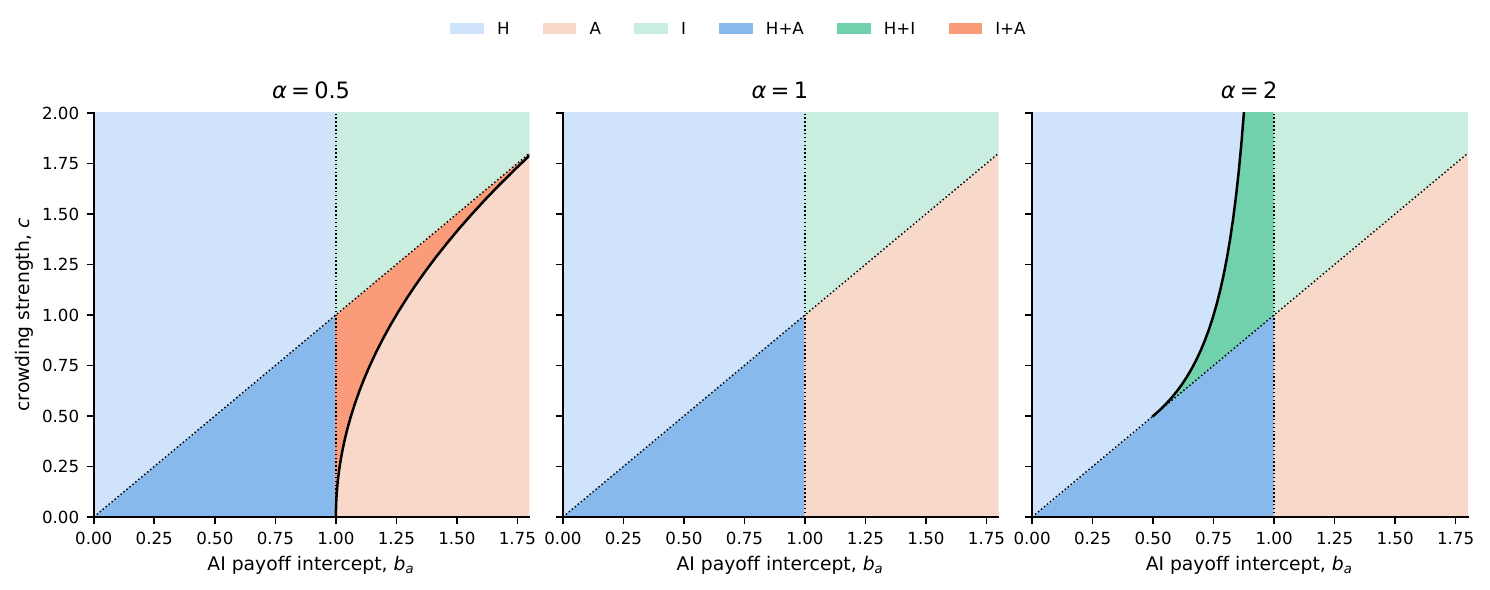}
\caption{\textbf{Attractors without spontaneous switching.}
Attractor sets in the $(b_a,c)$ plane for $m=0$, $b_h=1$, and $\eta=0$. $I$ denotes stable interior coexistence; combinations denote bistability.
The dotted lines are $b_a=b_h$ and $b_a=b_h\eta+c$; the solid line is $b_a=b_{a,f}(c)$ where the interior saddle-node exists. At $\alpha=1$ only $H+A$ bistability occurs. Superlinear crowding creates an $H+I$ region, whereas sublinear crowding creates an $I+A$ region.}
\label{supp:fig:SI-nonlinear-map}
\end{figure}

The phase portraits in Fig.~\ref{supp:fig:SI-nonlinear-portraits} show the corresponding basins when weak spontaneous switching moves the boundary equilibria into the interior. Under superlinear crowding, a transition out of autonomous work can end at partial, rather than near-complete, AI adoption. Under sublinear crowding, the low-adoption attractor is already a mixture of autonomous and AI-assisted work, but it can tip into near-complete reliance. The stable manifold of the saddle separates the initial conditions leading to the two outcomes.

\begin{figure}[t]
\centering
\includegraphics[width=.96\linewidth]
{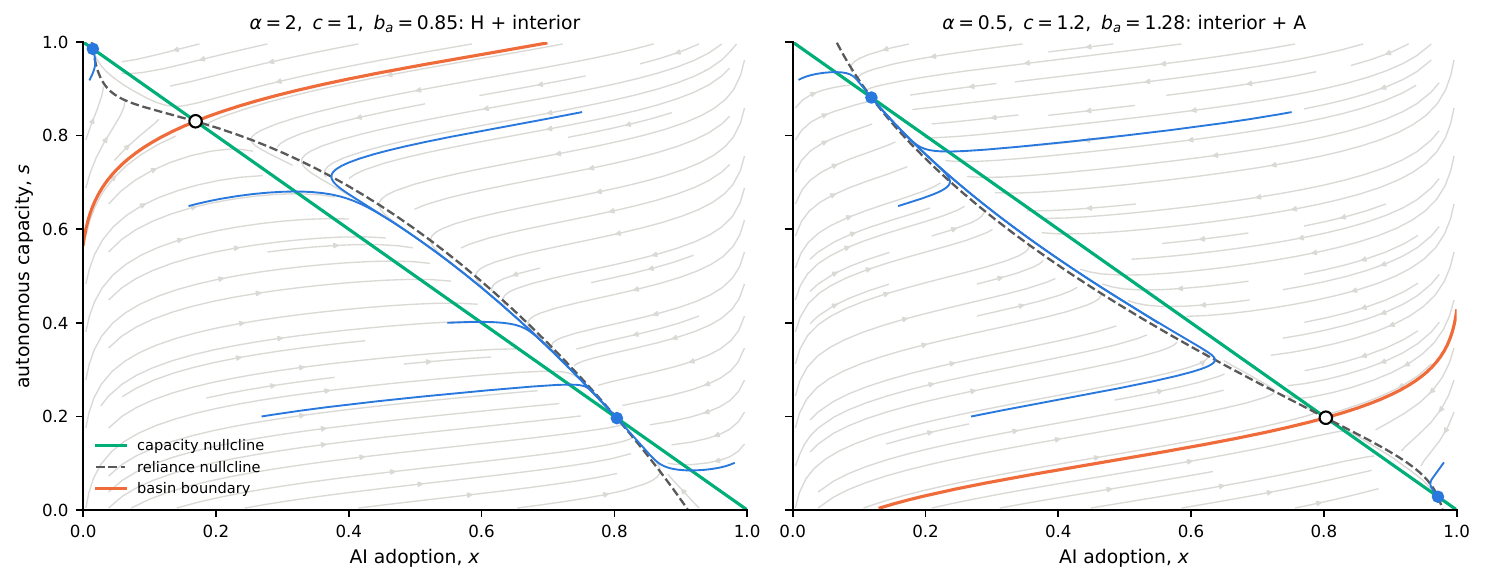}
\caption{\textbf{Phase portraits of the two additional bistable regimes.}
Left: $\alpha=2$, $c=1$, $b_a=0.85$ ($H+I$).
Right: $\alpha=1/2$, $c=1.2$, $b_a=1.28$ ($I+A$).
The green line is the skill nullcline, the gray dashed line the adoption nullcline, filled circles the stable equilibria, and the open circle the saddle. The orange curve is its stable manifold and blue curves are representative trajectories. Parameters: $b_h=1$, $\eta=0$, $\varepsilon=0.05$, $m=0.002$.}
\label{supp:fig:SI-nonlinear-portraits}
\end{figure}

\subsection{Spontaneous switching, folds and the cusp}

For $m>0$, every equilibrium is interior. Its adoption fraction parametrizes the equilibrium curve
\begin{equation}
b_a^{\rm eq}(x)=b_h-Bx+cx^\alpha
       +m\frac{2x-1}{x(1-x)}.
\label{supp:eq:ba-curve-alpha}
\end{equation}
Writing $\varphi(x)=x^{-2}+(1-x)^{-2}$ gives $(b_a^{\rm eq})'(x)=-\Gamma_\alpha(x)+m\varphi(x)$. As in the linear case, positive-slope branches are stable and negative-slope branches are saddles. Thus the bistability criterion is
\begin{equation}
\max_{0<x<1}\{\Gamma_\alpha(x)-m\varphi(x)\}>0.
\label{supp:eq:bistability-alpha}
\end{equation}
At $\alpha=1$, the maximum occurs at $x=1/2$ and this becomes $\Gamma>8m$. Figure~\ref{supp:fig:SI-nonlinear-branches} compares the equilibrium branches at the reference crowding strength and at $c=1$, where $\Gamma=0$. In the latter case the linear model is monostable, while nonlinear crowding can produce two distinct forms of hysteresis.

\begin{figure}[t]
\centering
\includegraphics[width=.96\linewidth]{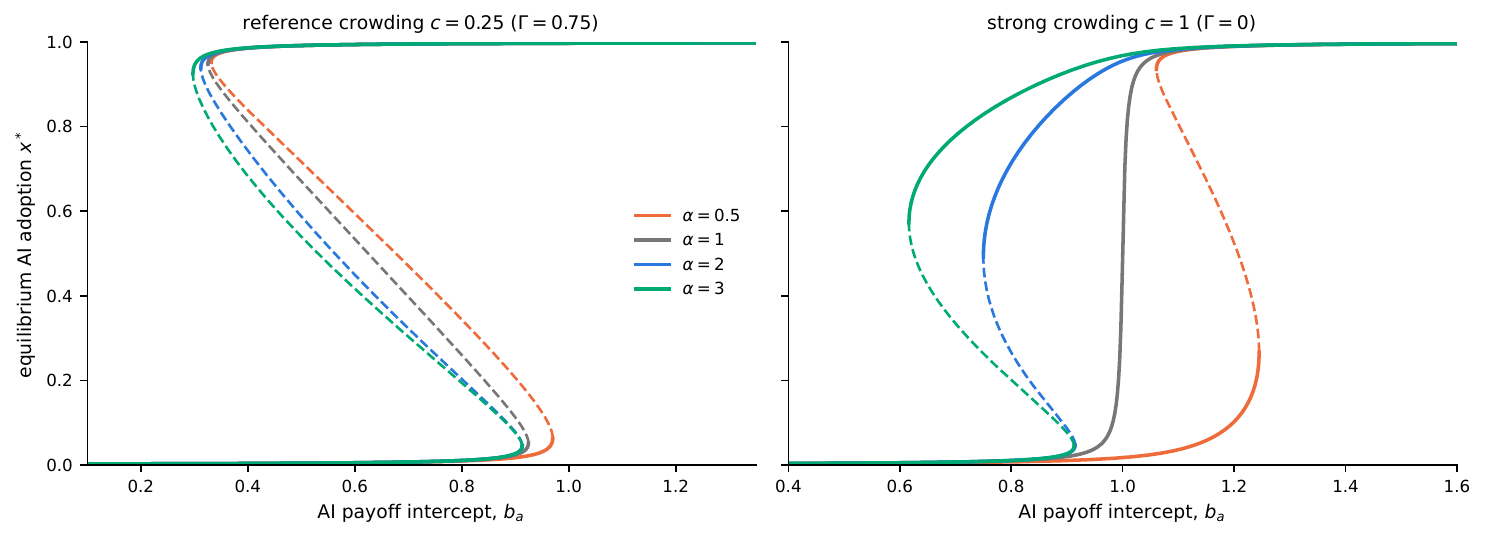}
\caption{\textbf{Equilibrium reliance under nonlinear crowding.}
$x^\star$ against $b_a$ at $m=0.002$, $b_h=1$, $\eta=0$ for $\alpha=1/2,1,2,3$. Solid branches are stable and dashed branches are saddles. Left: $c=0.25$; the familiar $H+A$ loop is deformed. Right: $c=1$; the linear model has no bistability, whereas $\alpha=2,3$ yield $H+I$ and $\alpha=1/2$ yields $I+A$.}
\label{supp:fig:SI-nonlinear-branches}
\end{figure}

The folds obey $\Gamma_\alpha(x)=m\varphi(x)$. Equivalently, the critical spontaneous-switching rate is
\begin{equation}
m_c(c,\eta,\alpha)=
\max\!\left\{0,\max_{0<x<1}
\frac{b_h(1-\eta)-c\alpha x^{\alpha-1}}{\varphi(x)}\right\}.
\label{supp:eq:nonlinear-critical-m}
\end{equation}
At a nondegenerate cusp, $x_c$ maximizes the ratio in Eq.~\eqref{supp:eq:nonlinear-critical-m}; therefore $\Gamma_\alpha'(x_c)\varphi(x_c)= \Gamma_\alpha(x_c)\varphi'(x_c)$. The linear model has $x_c=1/2$, whereas superlinear and sublinear crowding move the cusp toward lower and higher adoption, respectively. Figure~\ref{supp:fig:SI-nonlinear-cusp} shows this displacement and the corresponding $m_c(\eta)$. In the $m\to0$ limit, the retention threshold is $\eta_{\rm crit}=1-c\alpha/b_h$ for $\alpha\leq1$ (when it lies in the admissible interval), while for $\alpha>1$ bistability persists for some $m>0$ at every $\eta<1$ under global crowding.

\begin{figure}[t]
\centering
\includegraphics[width=.96\linewidth]{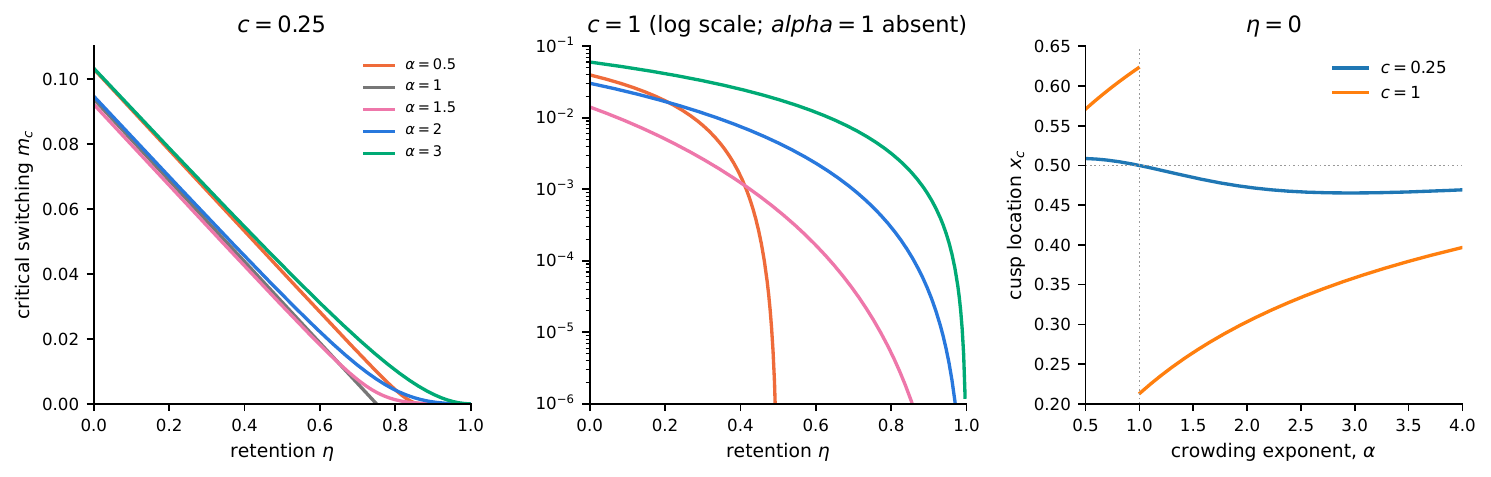}
\caption{\textbf{Critical switching rate and cusp location.}
Left and center: $m_c$ against retention $\eta$ for several crowding exponents at $c=0.25$ and $c=1$; bistability exists for $m<m_c$. The center panel uses a logarithmic vertical axis and omits $\alpha=1$, for which $m_c=0$ at $c=1$. Right: cusp adoption $x_c$ against $\alpha$ at $\eta=0$. At $c=1,\alpha=1$, $m_c=0$ and a positive-rate cusp is undefined, so the orange curve is interrupted. Other parameters: $b_h=1$.}
\label{supp:fig:SI-nonlinear-cusp}
\end{figure}

\subsection{Finite crowding samples}

The microscopic comparison rule uses the fraction $\hat x_k$ of AI users in a sample of $k$ agents. Under weak selection, averaging the sampled payoff gives the exact effective crowding function
\begin{equation}
f_k(x)=\E[\hat x_k^\alpha]
=\sum_{j=0}^{k}\binom{k}{j}x^j(1-x)^{k-j}
\left(\frac jk\right)^\alpha.
\label{supp:eq:sampled-nonlinear-crowding}
\end{equation}
For $\alpha=1$, $f_k(x)=x$ at every $k$. For $0<\alpha<1$, concavity implies $f_k(x)<x^\alpha$ in the interior, whereas for $\alpha>1$ convexity reverses the inequality. Figure~\ref{supp:fig:SI-nonlinear-sampling} shows both effects. In particular, finite sampling makes $f_k$ linear near $x=0$, regularizing the divergent derivative of $x^\alpha$ when $\alpha<1$. Equations~\eqref{supp:eq:nonlinear-critical-m} and the corresponding retention threshold use global crowding; for finite $k$, the exact finite-sample prediction replaces $x^\alpha$ by $f_k(x)$.

\begin{figure}[t]
\centering
\includegraphics[width=.96\linewidth]{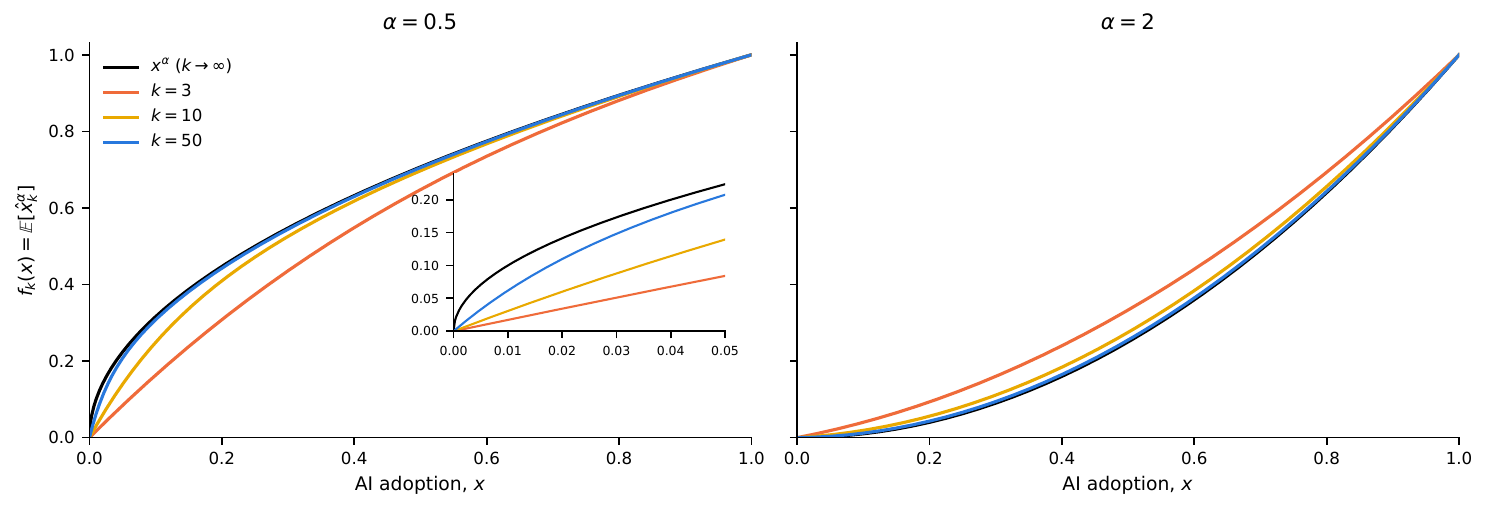}
\caption{\textbf{Finite-sample smoothing of nonlinear crowding.} $f_k(x)=\E[\hat x_k^\alpha]$ for $k=3,10,50$ compared with the global-crowding function $x^\alpha$ (black). Left: $\alpha=1/2$, with an inset near $x=0$; right: $\alpha=2$. For $\alpha=1$, $f_k(x)=x$ for all $k$ and is not shown.}
\label{supp:fig:SI-nonlinear-sampling}
\end{figure}

These calculations extend the reduced MFT. In the individual-skill model of Section~\ref{supp:sec:geff}, stationary payoff comparisons involve the mean skill of autonomous agents, $s_h$, rather than the population mean $s$. Accordingly, the effective local feedback is obtained by replacing $B$ with $b_h(1-\eta)/(1+\rho)$ in $\Gamma_\alpha(x)$ under the assumptions stated there. The reduced MFT curves above therefore establish possible regimes and mechanisms; their fold positions are not direct predictions for the finite-population agent-based model.

\begin{figure}[t]
\centering
\includegraphics[width=.96\linewidth]{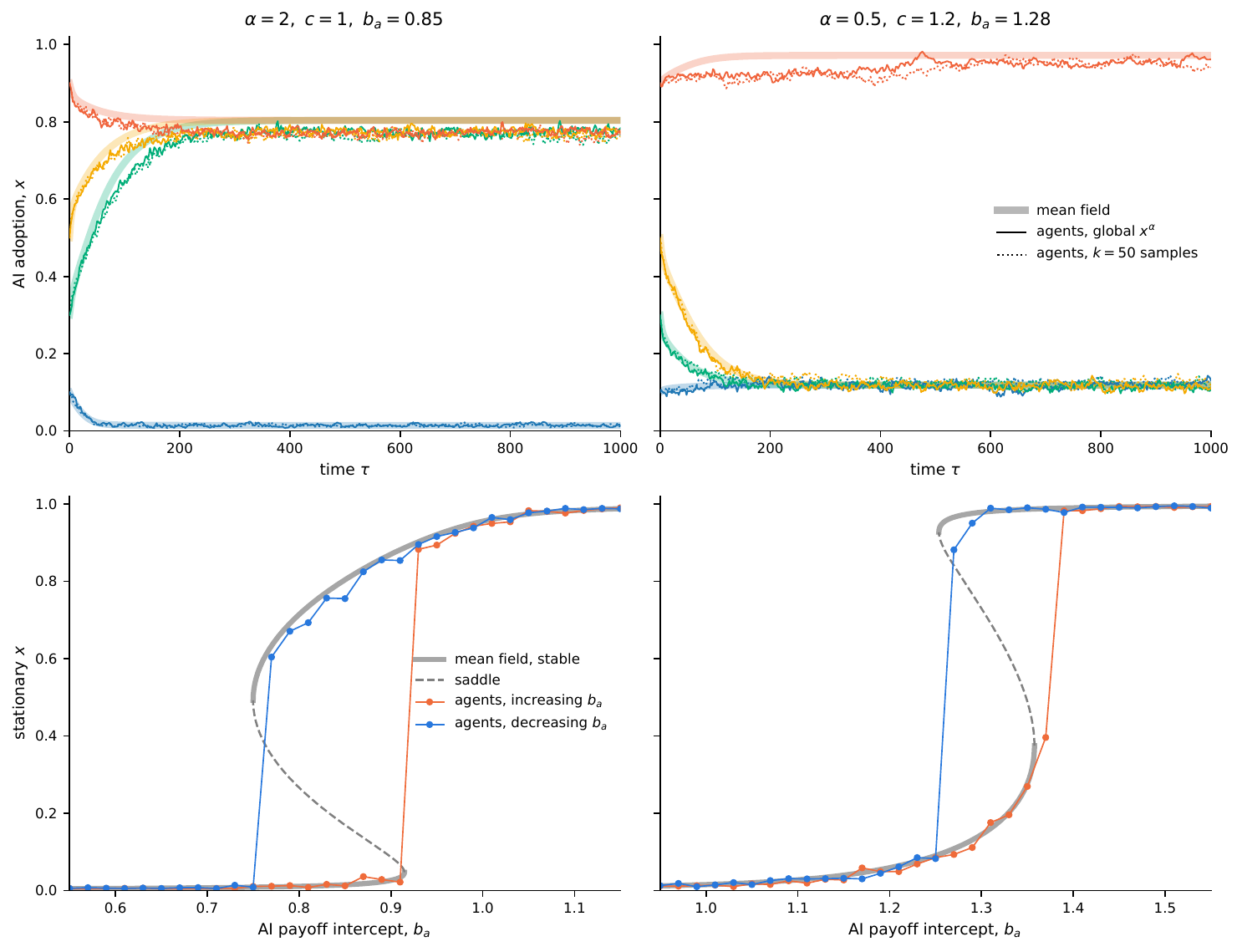}
\caption{\textbf{Agent-based check of nonlinear-crowding bistability.}
Top: AI adoption $x(\tau)$ from initial states
$(x_0,s_0)=(0.1,0.9),(0.3,0.7),(0.5,0.5),(0.9,0.1)$.
Thick translucent curves are reduced MFT trajectories; thin solid curves are agent-based runs with global crowding ($N=2\times10^4$), and dotted curves use $k=50$ sampled encounters.
Bottom: upward (orange) and downward (blue) agent-based sweeps,compared with stable (solid gray) and saddle (dashed gray) branches of the reduced MFT. The sweeps use $N=2000$, with 1500 relaxation and 300 measurement sweeps per value of $b_a$.
Left: $\alpha=2$, $c=1$, $b_a=0.85$ in the top panel, exhibiting autonomous/partial-reliance bistability. Right: $\alpha=1/2$, $c=1.2$, $b_a=1.28$ in the top panel, exhibiting low-coexistence/near-complete- reliance bistability. Other parameters are $b_h=1$, $\eta=0$, $\varepsilon=0.05$, $m=0.002$, and $\beta=0.5$.}
\label{supp:fig:SI-nonlinear-agents}
\end{figure}

\subsection{Agent-based check}

We next test whether the two additional regimes persist when agents carry individual skills and revise their strategies by Fermi comparison. In each revision a focal agent switches spontaneously with probability $\mu$ or compares its payoff with that of a uniformly drawn role model. We use $\beta=0.5$ and convert the reduced rates into simulation units according to Eq.~\eqref{supp:eq:units}:
\begin{equation}
w=\frac{\beta}{2},\qquad
\mu=\frac{mw}{1+mw},\qquad
\varepsilon_{\rm sw}=(1-\mu)w\varepsilon,\qquad
\tau=(1-\mu)w\,t_{\rm sw}.
\label{supp:eq:nonlinear-abm-units}
\end{equation}
Between revision events, each skill follows the exact exponential relaxation toward $1$ under autonomous work or $\eta$ under AI use. We compare both global crowding, $cx^\alpha$, and the payoff from $k=50$ sampled encounters, $c\hat{x}_{50}^{\alpha}$.

The upper panels of Fig.~\ref{supp:fig:SI-nonlinear-agents} compare trajectories from four initial adoption--skill states with the reduced MFT solutions. Both the autonomous/partial-reliance pair for $\alpha=2$ and the low-coexistence/near-complete-reliance pair for $\alpha=1/2$ remain observable. Finite sampling shifts the trajectories slightly, as expected from $f_k(x)\ne x^\alpha$. The lower panels use local upward and downward quasi-static sweeps. Their distinct paths verify history dependence in the agent model. Because payoffs depend on the skills of autonomous agents, and finite populations can switch basins before a deterministic fold, the simulated jumps need not coincide exactly with the reduced MFT fold positions.

\section{Effective feedback with individual skills}
\label{supp:sec:geff}

\subsection{Derivation}

Without the closure $\sigma=0$, the payoff difference entering Eq.~\eqref{supp:eq:replicator} contains $s_h$. At stationarity, the exact mean-skill relation [Eq.~\eqref{supp:eq:mean-skill-star}] gives 
\begin{equation}
    s^{\star}=1-(1-\eta)x^{\star},
\end{equation}
Under the skill-independent switching approximation, the stationary skill gap is
\begin{equation}
    s_h^{\star}-s_a^{\star}=(1-\eta)\frac{\rho}{1+\rho}.
\end{equation}
From Eq.~\eqref{supp:eq:decomposition}, $s=s_h-x(s_h-s_a)$, so
$s_h=s+x(s_h-s_a)$. Substituting,
\begin{align}
    s_h^{\star}
    &=1-(1-\eta)x^{\star}+x^{\star}(1-\eta)\frac{\rho}{1+\rho}
     =1-(1-\eta)x^{\star}\left[1-\frac{\rho}{1+\rho}\right]
     \nonumber\\
    &=1-(1-\eta)\frac{x^{\star}}{1+\rho}.
    \label{supp:eq:sh-star}
\end{align}
Autonomous agents are more skilled than average by $s_h^{\star}-s^{\star}=(1-\eta)x^{\star}\rho/(1+\rho)$. For $\alpha=1$, the
stationary payoff difference is then
\begin{align}
    b_a-cx^{\star}-b_hs_h^{\star}
    &=b_a-cx^{\star}-b_h+b_h(1-\eta)\frac{x^{\star}}{1+\rho}
    \nonumber\\
    &=b_a-b_h+\left[\frac{b_h(1-\eta)}{1+\rho}-c\right]x^{\star}
    \equiv b_a-b_h+\geff\,x^{\star}.
    \label{supp:eq:geff}
\end{align}
This has the form of the reduced MFT with
\begin{equation}
    \boxed{\;\geff=\frac{b_h(1-\eta)}{1+\rho}-c\;}
\end{equation}
in place of $\Gamma$ (\Meq{eq:gamma-eff}). The stationary states solve
\begin{equation}
    x(1-x)\left[b_a-b_h+\geff x\right]+m(1-2x)=0,
    \label{supp:eq:stationary-geff}
\end{equation}
For constant $\rho$, the stationary equation has the same algebraic form as the reduced MFT, so the results follow with $\Gamma\to\geff$: bistability requires $\geff>8m$, and for $m\ll\geff$ the folds are $b_{a,+}\simeq b_h-2\sqrt{m\geff}$ and $b_{a,-}\simeq b_h-\geff+2\sqrt{m\geff}$. 
When $\rho$ varies with adoption, the turning-point condition includes its derivative, as described below.

\paragraph{Limits.} For $\rho\to0$, $\geff\to\Gamma$ and the reduced MFT is recovered. For $\rho\to\infty$, $\geff\to-c$: every autonomous agent sits at $s=1$, the autonomous payoff no longer depends on adoption, and only crowding remains.

\subsection{Assumptions}
The MFT with covariance closure combines the weak-selection adoption equation with the skill-independent switching approximation for the stationary skill gap. It describes stationary self-consistency and does not provide a closed description of transient dynamics.
If $\rho$ depends on $x$, the equilibrium curve is
$$
b_a(x)=b_h-\frac{b_h(1-\eta)x}{1+\rho(x)}
       +cx^\alpha
       +m\frac{2x-1}{x(1-x)}.
$$
Its turning points satisfy
$$
0=
-\frac{b_h(1-\eta)}{1+\rho}
+\frac{b_h(1-\eta)x\rho'}{(1+\rho)^2}
+c\alpha x^{\alpha-1}
+m\left[\frac{1}{x^2}+\frac{1}{(1-x)^2}\right].
$$
For $\alpha=1$ and constant $\rho$, this reduces to the fold condition of the reduced MFT with $\Gamma\to\Gamma_{\mathrm{eff}}$.

\section{Stationary distribution of individual skills}
\label{supp:sec:distribution}

We work with the rescaled skill $u=(s-\eta)/(1-\eta)\in[0,1]$, which obeys $\dot u=\eps(1-u)$ under autonomous work and $\dot u=-\eps u$ under AI use.
Throughout this section rates are constant in time (stationary state) and, until Section~\ref{supp:sec:skill-dependent}, independent of individual skill.

\subsection{Kinetic equations}

Let $p_H(u,t)\,du$ be the fraction of the population that is autonomous with skill in $[u,u+du]$, and $p_A(u,t)\,du$ the same for AI users, with
\begin{equation}
    \int_0^1\left[p_H+p_A\right]du=1,\qquad x=\int_0^1p_A\,du .
\end{equation}
Consider autonomous agents with skill in $[u,u+du]$. Their number changes for three reasons.
\begin{enumerate}
\item \emph{Drift of skill.} Skill moves with velocity $v_H(u)=\eps(1-u)$.
Agents enter the bin through its left edge at rate $v_H(u)p_H(u)$ and leave through its right edge at rate $v_H(u+du)p_H(u+du)$; the net change is $-\partial_u[v_Hp_H]\,du$.
\item \emph{Departures.} Each autonomous agent switches to AI at rate $r_{+}$:
loss $r_{+}p_H\,du$.
\item \emph{Arrivals.} AI users with skill in the bin switch back at rate $r_{-}$, keeping their skill: gain $r_{-}p_A\,du$.
\end{enumerate}
For AI users the velocity is $v_A(u)=-\eps u$. Hence
\begin{align}
    \partial_tp_H&=-\partial_u\left[\eps(1-u)\,p_H\right]-r_{+}p_H+r_{-}p_A,
    \label{supp:eq:kinH}\\
    \partial_tp_A&=+\partial_u\left[\eps u\,p_A\right]+r_{+}p_H-r_{-}p_A .
    \label{supp:eq:kinA}
\end{align}
These are the transport equations of a variable driven by dichotomous noise \cite{werner_horsthemke_noise-induced_1984, bena_dichotomous_2006}.

\paragraph{Checks.} Integrating Eq.~\eqref{supp:eq:kinA} over $u$ gives $\dot x=[\eps up_A]_0^1+r_{+}(1-x)-r_{-}x$; the boundary term vanishes (Section~\ref{supp:sec:zero-flux}), recovering Eq.~\eqref{supp:eq:x-rates}. Multiplying the sum of Eqs.~\eqref{supp:eq:kinH}--\eqref{supp:eq:kinA} by $u$, integrating, and integrating the transport terms by parts,
\begin{equation}
    \frac{d\bar u}{dt}=\int_0^1\eps(1-u)p_H\,du-\int_0^1\eps u\,p_A\,du
    =\eps\left[(1-x)-\bar u\right],
\end{equation}
since $\int(1-u)p_H=(1-x)-\int up_H$ and $\int up_H+\int up_A=\bar u$. This is Eq.~\eqref{supp:eq:mean-skill} in rescaled units.

\subsection{Zero flux}
\label{supp:sec:zero-flux}

Adding Eqs.~\eqref{supp:eq:kinH} and \eqref{supp:eq:kinA}, the switching terms cancel in pairs, and the total density $p=p_H+p_A$ obeys a continuity equation,
\begin{equation}
    \partial_tp+\partial_uJ=0,\qquad
    J(u)=\eps(1-u)\,p_H(u)-\eps u\,p_A(u),
\end{equation}
where $J(u)$ is the net number of agents per unit time whose skill crosses the value $u$ upward. At stationarity $\partial_uJ=0$, so $J$ is constant. Both velocities point into the interval: $v_H\ge0$ with $v_H(1)=0$, and $v_A\le0$ with $v_A(0)=0$. No probability can leave $[0,1]$, so $J(0)=J(1)=0$ and hence $J\equiv0$:
\begin{equation}
    \boxed{\;u\,p_A(u)=(1-u)\,p_H(u)\equiv f(u).\;}
    \label{supp:eq:zero-flux}
\end{equation}
At each skill level the upward drift of autonomous agents balances the downward drift of AI users.

\subsection{Solution}

Using Eq.~\eqref{supp:eq:zero-flux}, $p_H=f/(1-u)$ and $p_A=f/u$. The stationary form of Eq.~\eqref{supp:eq:kinH} becomes
\begin{equation}
    0=-\eps f'(u)-r_{+}\frac{f}{1-u}+r_{-}\frac{f}{u},
    \qquad\text{i.e.}\qquad
    \frac{f'}{f}=\frac{r_{-}}{\eps}\,\frac1u-\frac{r_{+}}{\eps}\,\frac1{1-u}.
    \label{supp:eq:f-ode}
\end{equation}
With $a=r_{-}/\eps$ and $b=r_{+}/\eps$, and since $\frac{d}{du}\ln u=1/u$ and $\frac{d}{du}\ln(1-u)=-1/(1-u)$, Eq.~\eqref{supp:eq:f-ode} reads $\frac{d}{du}\ln f=\frac{d}{du}[a\ln u+b\ln(1-u)]$, so
\begin{equation}
    f(u)=K\,u^{a}(1-u)^{b}.
\end{equation}
Therefore
\begin{align}
    p_H(u) &=K\,u^{a}(1-u)^{b-1},\\
    p_A(u) &=K\,u^{a-1}(1-u)^{b}, \\
    p(u) &= K\,u^{a-1}(1-u)^{b-1}\left[u+(1-u)\right]=K\,u^{a-1}(1-u)^{b-1}.
\end{align}
With the Beta function $B(a,b)=\int_0^1u^{a-1}(1-u)^{b-1}du$, normalization gives $K=1/B(a,b)$:
\begin{equation}
    \boxed{\;p(u)=\frac{u^{a-1}(1-u)^{b-1}}{B(a,b)},\qquad
    u\sim\mathrm{Beta}\!\left(\frac{r_{-}}{\eps},\frac{r_{+}}{\eps}\right).\;}
    \label{supp:eq:beta}
\end{equation}

\subsection{Fraction of AI users and parametrization by $\rho$}

Integrating $p_A$ gives $x=B(a,b+1)/B(a,b)$. This ratio follows from two properties of the Beta function. First, since $u+(1-u)=1$,
\begin{equation}
    B(a,b)=\int_0^1u^{a-1}(1-u)^{b-1}\left[u+(1-u)\right]du=B(a+1,b)+B(a,b+1).
\end{equation}
Second, integrating by parts,
\begin{equation}
    B(a,b+1)=\int_0^1u^{a-1}(1-u)^{b}\,du
    =\left[\frac{u^{a}}{a}(1-u)^{b}\right]_0^1+\frac{b}{a}\int_0^1u^{a}(1-u)^{b-1}\,du
    =\frac{b}{a}\,B(a+1,b).
\end{equation}
Substituting $B(a+1,b)=(a/b)\,B(a,b+1)$ into the first relation gives $B(a,b)=B(a,b+1)\,(a+b)/b$, hence
\begin{equation}
    x=\frac{B(a,b+1)}{B(a,b)}=\frac{b}{a+b}=\frac{r_{+}}{r_{+}+r_{-}},
    \label{supp:eq:x-beta}
\end{equation}
which is the balance of flows $r_{+}(1-x)=r_{-}x$; it is therefore not assumed but follows from the solution. With $\lambda=r_{+}+r_{-}$ and $\rho=\eps/\lambda$, $r_{+}=\lambda x$ and $r_{-}=\lambda(1-x)$, so
\begin{equation}
    u\sim\mathrm{Beta}\!\left(\frac{1-x^{\star}}{\rho},\frac{x^{\star}}{\rho}\right),
    \label{supp:eq:beta-rho}
\end{equation}
which is \Meq{eq:beta-law}. Its mean is $a/(a+b)=1-x^{\star}$, consistent with Eq.~\eqref{supp:eq:mean-skill-star}.

\subsection{Skills within each strategy}

From the solution, the conditional densities are $p_H/(1-x)\propto u^{a}(1-u)^{b-1}$ and $p_A/x\propto u^{a-1}(1-u)^{b}$:
\begin{equation}
    u\mid H\sim\mathrm{Beta}(a+1,b),\qquad u\mid A\sim\mathrm{Beta}(a,b+1).
\end{equation}
Their means are
\begin{equation}
    u_h=\frac{a+1}{a+b+1},\qquad u_a=\frac{a}{a+b+1},\qquad
    u_h-u_a=\frac{1}{a+b+1}=\frac{\eps}{\eps+\lambda}=\frac{\rho}{1+\rho},
    \label{supp:eq:u-means}
\end{equation}
which reproduces Eq.~\eqref{supp:eq:gap} exactly: for skill-blind switching the covariance result is exact. Using $a+1=(r_{-}+\eps)/\eps$, $a+b+1=(\lambda+\eps)/\eps$ and $r_{-}=\lambda(1-x^{\star})$,
\begin{equation}
    u_h=\frac{\lambda(1-x^{\star})+\eps}{\lambda+\eps}
    =1-\frac{\lambda x^{\star}}{\lambda+\eps}=1-\frac{x^{\star}}{1+\rho},
\end{equation}
which is Eq.~\eqref{supp:eq:sh-star} in rescaled units.

\subsection{Shape of the distribution}

Near the endpoints $p(u)\sim u^{a-1}$ and $p(u)\sim(1-u)^{b-1}$. The density diverges at $u=0$ if $a<1$, i.e.\ $\rho>1-x^{\star}$, and at $u=1$ if $b<1$, i.e.\ $\rho>x^{\star}$. If $a,b>1$ the distribution is unimodal with mode $(a-1)/(a+b-2)$ and variance
\begin{equation}
    \mathrm{Var}[u]=\frac{ab}{(a+b)^2(a+b+1)}=\frac{x^{\star}(1-x^{\star})\,\rho}{1+\rho},
    \label{supp:eq:u-variance}
\end{equation}
which increases with $\rho$. For $x^{\star}=1/2$, the law is $\mathrm{Beta}(1/2\rho,1/2\rho)$: unimodal for $\rho<1/2$, uniform at $\rho=1/2$, and U-shaped for $\rho>1/2$. This change of shape is the analog of the noise-induced transitions studied for dichotomous noise
\cite{sancho_stochastic_1984, werner_horsthemke_noise-induced_1984}.

\subsection{Cross-check through the moments}

The same law follows from the moments. Let $M_n=\E[u^n]$ and $Q_n=\E[\chi u^n]$ with $\chi=\1(z=H)$, so that $\dot u=\eps(\chi-u)$. At stationarity, $0=\frac{d}{dt}\E[u^n]=n\eps\,\E[u^{n-1}(\chi-u)]$, hence
\begin{equation}
    Q_{n-1}=M_n .
    \label{supp:eq:moment-identity}
\end{equation}
The stationarity of $\E[\chi u^{n-1}]$ gives, from learning, switching away and switching in (using $\chi^2=\chi$),
\begin{align}
    0&=(n-1)\eps\,\E[\chi u^{n-2}(1-u)]-r_{+}\E[\chi u^{n-1}]+r_{-}\E[(1-\chi)u^{n-1}] \\
    &=(n-1)\eps(Q_{n-2}-Q_{n-1})-r_{+}Q_{n-1}+r_{-}(M_{n-1}-Q_{n-1}).
\end{align}
Eliminating the $Q$'s with Eq.~\eqref{supp:eq:moment-identity},
\begin{equation}
    M_n=\frac{r_{-}+(n-1)\eps}{r_{+}+r_{-}+(n-1)\eps}\,M_{n-1}
    =\prod_{j=0}^{n-1}\frac{a+j}{a+b+j},
\end{equation}
the moments of $\mathrm{Beta}(a,b)$. A distribution on $[0,1]$ is determined by its moments, so the two derivations agree. Multiplying the zero-flux condition Eq.~\eqref{supp:eq:zero-flux} by $u^{n-1}$ and integrating gives $\E[(1-\chi)u^{n}]=\E[\chi u^{n-1}]-\E[\chi u^{n}]$, i.e. $M_n-Q_n=Q_{n-1}-Q_n$, which is Eq.~\eqref{supp:eq:moment-identity}: the moment identity is the integrated form of the zero-flux condition.

\paragraph{Exact stationary variance--gap identity.}
Eq.~\eqref{supp:eq:moment-identity} follows from the skill dynamics alone: strategy changes leave $u$ unchanged, so it holds for any switching rule, including the skill-dependent rates of Section~\ref{supp:sec:skill-dependent}. For $n=1$ and $n=2$ it gives
$$
M_1 = Q_0 = 1-x^\star, \qquad Q_1 = M_2 ,
$$
that is, $\mathbb{E}[u]=\mathbb{E}[\chi]=1-x^\star$ and $\mathbb{E}[\chi u]=\mathbb{E}[u^2]$. Hence
$$
\operatorname{Cov}[\chi,u] = Q_1 - Q_0 M_1 = M_2 - M_1^2 = \operatorname{Var}[u]:
$$
at stationarity, the covariance between autonomous work and skill equals the variance of skills. Since $\mathbf{1}(z=A)=1-\chi$ and $s=\eta+(1-\eta)u$, the strategy--skill covariance of Eq.~\eqref{supp:eq:first-sigma} is
$$
\sigma^\star = -(1-\eta)\operatorname{Cov}[\chi,u]
= -(1-\eta)\operatorname{Var}[u]
= -\frac{\operatorname{Var}[s]}{1-\eta}.
$$
Using $\sigma = x(1-x)(s_a-s_h)$, the skill gap is
$$
s_h^\star - s_a^\star = \frac{\operatorname{Var}[s]}{(1-\eta)\,x^\star(1-x^\star)},
\qquad\text{equivalently}\qquad
u_h - u_a = \frac{\operatorname{Var}[u]}{x^\star(1-x^\star)} .
$$
For skill-independent switching, inserting the Beta variance of Eq.~\eqref{supp:eq:u-variance} recovers $u_h-u_a=\rho/(1+\rho)$, Eq.~\eqref{supp:eq:u-means}.

\subsection{Skill-dependent switching}
\label{supp:sec:skill-dependent}

In the payoff-driven model, $r_{+}$ depends on the agent's own skill [Eq.~\eqref{supp:eq:rplus-skill}], while $r_{-}$ does not. Eqs.~\eqref{supp:eq:kinH}--\eqref{supp:eq:kinA} hold with $r_{+}\to r_{+}(u)$, and the zero-flux argument is unchanged. Eq.~\eqref{supp:eq:f-ode} becomes
\begin{equation}
    \frac{f'}{f}=\frac{r_{-}}{\eps}\,\frac1u-\frac{r_{+}(u)}{\eps}\,\frac1{1-u}.
\end{equation}
Writing $r_{+}(v)=r_{+}(1)+[r_{+}(v)-r_{+}(1)]$ and integrating from $0$ to $u$,
\begin{equation}
    \ln f(u)=\text{const}+\frac{r_{-}}{\eps}\ln u+\frac{r_{+}(1)}{\eps}\ln(1-u)
    -\frac1\eps\int_0^u\frac{r_{+}(v)-r_{+}(1)}{1-v}\,dv,
\end{equation}
where the last integral is finite for smooth $r_{+}$. Hence
\begin{equation}
    \boxed{\;p(u)=K\,u^{r_{-}/\eps-1}(1-u)^{r_{+}(1)/\eps-1}\,
    \exp\!\left[-\frac1\eps\int_0^u\frac{r_{+}(v)-r_{+}(1)}{1-v}\,dv\right].\;}
    \label{supp:eq:general-density}
\end{equation}
The rates depend on $x$ and on the skills of autonomous role models, so Eq.~\eqref{supp:eq:general-density} must be complemented by the self-consistency conditions $x=\int_0^1p_A\,du$ and $r_{-}=\mu+(1-\mu)(1-x)\langle\phi\rangle_H$, solved numerically.

\paragraph{First correction at weak selection.} At first order in $\beta$, Eq.~\eqref{supp:eq:rplus-skill} gives an affine rate. In rescaled skill, $s-s_h=(1-\eta)(u-u_h)$, so
\begin{equation}
    r_{+}(u)=A_0+A_1u,\qquad
    A_1=-(1-\mu)\,x\,\frac{\beta}{4}\,b_h(1-\eta)<0 \quad(\text{per sweep}).
\end{equation}
Then $r_{+}(v)-r_{+}(1)=A_1(v-1)$, the integrand is $-A_1$, and the exponential factor is $e^{A_1u/\eps}$:
\begin{equation}
    p(u)\propto u^{a-1}(1-u)^{b'-1}e^{A_1u/\eps},\qquad
    a=\frac{r_{-}}{\eps},\quad b'=\frac{A_0+A_1}{\eps},
\end{equation}
with rates and $\eps$ in the same units. Relative to the Beta law, the skill dependence changes the exponent at $u=1$ and multiplies the density by an exponential factor.

\subsection{Exact skill gap for skill-dependent switching}

Multiply the stationary Eq.~\eqref{supp:eq:kinH} by $u$ and integrate. The transport term gives, after integration by parts,
\begin{equation}
    -\int_0^1u\,\partial_u[\eps(1-u)p_H]\,du
    =-\left[u\,\eps(1-u)p_H\right]_0^1+\eps\int_0^1(1-u)p_H\,du
    =\eps(1-x)(1-u_h),
\end{equation}
where $u_h$ is the mean rescaled skill of autonomous agents. The switching terms give $-\int ur_{+}(u)p_H\,du+r_{-}\int up_A\,du =-(1-x)\langle ur_{+}\rangle_H+r_{-}xu_a$. Hence
\begin{equation}
    0=\eps(1-x)(1-u_h)-(1-x)\left[\langle r_{+}\rangle_Hu_h+\mathrm{Cov}_H(r_{+},u)\right]+r_{-}xu_a .
\end{equation}
where $\langle\cdot\rangle_H$ denotes an average over autonomous agents, whose skills are distributed as $p_H(u)/(1-x)$, and
\begin{equation}
    \mathrm{Cov}_H(r_{+},u)=\langle r_{+}u\rangle_H-\langle r_{+}\rangle_H\,u_h
    =\frac{1}{1-x}\int_0^1\left[r_{+}(u)-\langle r_{+}\rangle_H\right]
    \left(u-u_h\right)p_H(u)\,du
\end{equation}
is the covariance, among autonomous agents, between the rate of switching to AI and skill. It vanishes for skill-blind switching and is negative when more skilled agents are less likely to adopt AI.
With the balance of flows $r_{-}x=\langle r_{+}\rangle_H(1-x)$, dividing by $1-x$,
\begin{equation}
    \eps(1-u_h)=\langle r_{+}\rangle_H(u_h-u_a)+\mathrm{Cov}_H(r_{+},u).
    \label{supp:eq:gap-general-1}
\end{equation}
The exact mean-skill relation $\bar u=(1-x)u_h+xu_a=1-x$ gives $1-u_h=x(1-\Delta)$ with $\Delta=u_h-u_a$. Substituting into Eq.~\eqref{supp:eq:gap-general-1} and using $\langle r_{+}\rangle_H=\bar\lambda x$, with $\bar\lambda=\langle r_{+}\rangle_H+r_{-}$,
\begin{equation}
    \eps x(1-\Delta)=\bar\lambda x\,\Delta+\mathrm{Cov}_H(r_{+},u)
    \;\Longrightarrow\;
    \boxed{\;u_h-u_a=\frac{\eps-\mathrm{Cov}_H(r_{+},u)/x}{\eps+\bar\lambda}.\;}
\end{equation}
Since more skilled autonomous agents adopt AI less often,
$\mathrm{Cov}_H(r_{+},u)<0$, and the skill gap is larger than $\rho/(1+\rho)$. At first order in $\beta$, $\mathrm{Cov}_H(r_{+},u)=A_1\mathrm{Var}_H(u)$. This is the origin of the excess of highly skilled agents in \Mfig{fig:agent_capacity_distributions}, panel B, and of the residual deviations of the simulated fraction of AI users from the MFT with covariance closure prediction.

\subsection{Predicting the distribution from the model parameters}

The Beta parameters in Eq.~\eqref{supp:eq:beta-rho} follow from the model: $\rho$ from Eq.~\eqref{supp:eq:lambda-star} and $x^{\star}$ from Eq.~\eqref{supp:eq:stationary-geff}. Since $\lambda^{\star}$ depends weakly on $x^{\star}$, the two are obtained by iteration: start from a guess for $x$, compute $\lambda^{\star}$, $\rho$ and $\geff$, solve Eq.~\eqref{supp:eq:stationary-geff} for $x$, and repeat. Table~\ref{supp:tab:beta-example} gives the values for the parameters of \Mfig{fig:agent_capacity_distributions}. Because $|\Gamma|=0.2$ is small at these parameters, the correction $b_h(1-\eta)\rho/(1+\rho)$ is comparable to it even for $\rho=0.03$, and the reduced MFT overestimates the fraction of AI users, increasingly so as $\rho$ grows.

\begin{table}[h]
\centering
\caption{Predicted parameters of the skill distribution for
\Mfig{fig:agent_capacity_distributions} ($b_h=1$, $c=1.2$, $\eta=0$, $\beta=1$, $m=0.002$), in rescaled units. The last column gives the prediction of the reduced MFT, $\Gamma=b_h(1-\eta)-c=-0.2$ in place of $\geff$; the agent-based values are $0.382$, $0.389$ and $0.395$.}
\label{supp:tab:beta-example}
\small
\begin{tabular}{@{}cccccccc@{}}
\toprule
$\eps$ & $b_a$ & $\lambda^{\star}$ & $\rho$ & $\geff$ & $x^{\star}$ & Beta parameters & $x^{\star}$ with $\sigma=0$\\
\midrule
$0.03$ & $1.090$ & $1.004$ & $0.030$ & $-0.229$ & $0.400$ & $(20.1,\,13.4)$ & $0.454$\\
$0.30$ & $1.170$ & $1.004$ & $0.299$ & $-0.430$ & $0.399$ & $(2.01,\,1.34)$ & $0.810$\\
$3.0$  & $1.378$ & $1.004$ & $2.99$  & $-0.949$ & $0.400$ & $(0.201,\,0.134)$ & $0.989$\\
\bottomrule
\end{tabular}
\end{table}

\section{Crowding from diversity of AI outputs}
\label{supp:sec:urn}

\subsection{Output distribution and originality costs}

Every output falls into one of $M$ idea categories. Autonomous outputs are independent draws from $\mathbf p=(p_1,\dots,p_M)$ and AI-assisted outputs from $\mathbf q=(q_1,\dots,q_M)$. Each output is compared with the output of one randomly chosen competitor, and a coincidence of categories costs its author $\omega\ge0$.

\paragraph{Coincidence probabilities.} Two independent outputs fall in category $j$ with the product of their probabilities, so, summing over $j$,
\begin{equation}
    P_{HH}=\sum_jp_j^2,\qquad P_{HA}=\sum_jp_jq_j,\qquad P_{AA}=\sum_jq_j^2 .
\end{equation}

\paragraph{Expected costs.} The competitor uses AI with probability $x$. Conditioning on its strategy,
\begin{equation}
    K_H=\omega\left[(1-x)P_{HH}+xP_{HA}\right],\qquad
    K_A=\omega\left[(1-x)P_{HA}+xP_{AA}\right].
\end{equation}
Only payoff differences matter in Eq.~\eqref{supp:eq:fermi}, so the cost borne by autonomous agents can be moved into the AI payoff:
\begin{align}
    K_A-K_H
    &=\omega\left[(1-x)(P_{HA}-P_{HH})+x(P_{AA}-P_{HA})\right]
    \nonumber\\
    &=\omega(P_{HA}-P_{HH})+\omega\,x\left[P_{AA}-P_{HA}-P_{HA}+P_{HH}\right]
    \nonumber\\
    &=\omega(P_{HA}-P_{HH})+\omega\,x\,\|\mathbf q-\mathbf p\|^2,
\end{align}
since $\|\mathbf q-\mathbf p\|^2=\sum_j(q_j^2-2q_jp_j+p_j^2)=P_{AA}-2P_{HA}+P_{HH}$.
Comparing with $\pi_A=b_a-cx$,
\begin{equation}
    c=\omega\|\mathbf q-\mathbf p\|^2\ge0,\qquad
    b_a\to b_a-\omega(P_{HA}-P_{HH}).
\end{equation}
This is \Meqs{eq:c-general}.

\paragraph{Uniform benchmark.} For $p_j=1/M$, $P_{HH}=\sum_jM^{-2}=1/M$ and $P_{HA}=M^{-1}\sum_jq_j=1/M$ for any $\mathbf q$, so the intercept shift vanishes and
\begin{equation}
    c=\omega d,\qquad d=\sum_jq_j^2-\frac1M .
\end{equation}
Expanding the square and using $\sum_jq_j=1$,
\begin{equation}
    \sum_j\left(q_j-\frac1M\right)^2=\sum_jq_j^2-\frac2M\sum_jq_j+\frac{M}{M^2}
    =\sum_jq_j^2-\frac1M=d\ge0,
\end{equation}
with equality only for uniform $\mathbf q$. With the effective number of categories $M_{\mathrm{eff}}=1/\sum_jq_j^2$ (inverse Simpson index \cite{simpson_measurement_1949}), $d=1/M_{\mathrm{eff}}-1/M$. The uniform benchmark is a reference case of maximal human diversity; the general results above hold for any $\mathbf p$.

\paragraph{Homogenization of the population.} Two randomly chosen outputs are both autonomous with probability $(1-x)^2$, mixed with probability $2x(1-x)$, and both AI-assisted with probability $x^2$. Their coincidence probability is $P(x)=(1-x)^2P_{HH}+2x(1-x)P_{HA}+x^2P_{AA}$. Relative to a fully autonomous population,
\begin{align}
    R(x)=P(x)-P_{HH}
    &=(-2x+x^2)P_{HH}+(2x-2x^2)P_{HA}+x^2P_{AA}
    \nonumber\\
    &=2x(P_{HA}-P_{HH})+x^2\left(P_{AA}-2P_{HA}+P_{HH}\right),
\end{align}
which for the uniform benchmark reduces to $R=d\,x^2$.

\subsection{The shape of crowding}

Suppose an output is compared with $k$ independently sampled competitors and incurs an originality penalty $\omega>0$ if it coincides with at least $r$ of them, where $1\le r\le k$. For an output in category $j$, each competitor coincides with probability $\pi_j(x)=(1-x)p_j+xq_j$. The number of coincidences therefore follows $\mathrm{Bin}(k,\pi_j(x))$, and the expected penalty is
\begin{equation}
    \omega P_{\ge r}(\pi_j(x)),
    \qquad
    P_{\ge r}(\pi)
    =\sum_{n=r}^{k}\binom{k}{n}\pi^n(1-\pi)^{k-n}.
\end{equation}

Averaging over the output categories gives the penalties for AI-assisted and autonomous work,
$$
    K_A(x)=\omega\sum_j q_jP_{\ge r}(\pi_j(x)),
    \qquad
    K_H(x)=\omega\sum_j p_jP_{\ge r}(\pi_j(x)).
$$
The quantity entering the payoff difference is their difference.
Writing $\delta_j=q_j-p_j$, we obtain
\begin{equation}
    G(x)=K_A(x)-K_H(x)
    =\omega\sum_j\delta_jP_{\ge r}(p_j+x\delta_j).
\end{equation}
The baseline difference $G(0)$ can be absorbed into $b_a$, leaving the adoption-dependent relative penalty $G(x)-G(0)$.

For a single competitor, $k=r=1$, we have
$P_{\ge1}(\pi)=\pi$, so
$$
    G(x)-G(0)
    =\omega x\sum_j\delta_j^2
    =\omega x\|\mathbf q-\mathbf p\|^2.
$$
This recovers the linear crowding coefficient
$c=\omega\|\mathbf q-\mathbf p\|^2$. 
For multiple competitors, the relative penalty remains
non decreasing:
\begin{equation}
    G'(x)=\omega\sum_j\delta_j^2
    P_{\ge r}'(p_j+x\delta_j)\ge0,
\end{equation}
because the binomial-tail probability increases with the coincidence probability. Consequently, $G(x)-G(0)\ge0$ for any output distributions $\mathbf p$ and $\mathbf q$. For uniform $\mathbf p$, $G(0)=0$, so $G(x)$ itself is nonnegative.
Its curvature, however, is not fixed:
\begin{equation}
    G''(x)=\omega\sum_j\delta_j^3
    P_{\ge r}''(p_j+x\delta_j).
\end{equation}
For $r=1$ and $k>1$,
$P_{\ge1}(\pi)=1-(1-\pi)^k$ is concave, making both $K_A(x)$ and $K_H(x)$ concave. Their difference need not be concave, because the coefficients $\delta_j^3$ can have either sign.

Multiple-competitor overlap therefore provides a microscopic source of nonlinear relative crowding. The power law $cx^\alpha$ used in the main text is a phenomenological representation of this response: $\alpha<1$ describes concave crowding, whereas $\alpha>1$ describes convex crowding.

\subsection{Diversity of AI outputs}

With $c=\omega d$, $\geff=b_h(1-\eta)/(1+\rho)-\omega d$. The results of Section~\ref{supp:sec:meanfield} then give:
\begin{itemize}
\item bistability requires $\geff>8m$, i.e.
\begin{equation}
    d<d_c=\frac1\omega\left[\frac{b_h(1-\eta)}{1+\rho}-8m\right];
\end{equation}
\item for $m\ll\geff$, the folds are $b_{a,+}\simeq b_h-2\sqrt{m\geff}$ and $b_{a,-}\simeq b_h-\geff+2\sqrt{m\geff}$. Their derivatives with respect to $d$ are
\begin{equation}
    \frac{\partial b_{a,+}}{\partial d}=\omega\sqrt{\frac m{\geff}},\qquad
    \frac{\partial b_{a,-}}{\partial d}=\omega\left(1-\sqrt{\frac m{\geff}}\right),
\end{equation}
using $\partial\geff/\partial d=-\omega$. For $m\ll\geff$, reducing $d$ lowers the lower fold almost one-for-one (in units of $\omega$) and barely moves the upper fold;
\item bistability disappears for $\eta>\eta_{\mathrm{crit}}$, obtained by solving $b_h(1-\eta)/(1+\rho)=\omega d+8m$:
\begin{equation}
    \eta_{\mathrm{crit}}=1-\frac{(1+\rho)(\omega d+8m)}{b_h}.
\end{equation}
\end{itemize}

\section{Simulation details and finite-size test}
\label{supp:sec:finite}

\paragraph{Quasi-static continuation (\Mfig{fig:agent-hysteresis}).} $b_a$ takes
$61$ equally spaced values in $[0.15,1.10]$. At each value the population
relaxes for $1200$ sweeps and is measured every $5$ sweeps for $300$ further
sweeps. The upward sweep starts from $x=0.02$ with $s_i=1$; the downward sweep
starts from the final state of the upward one. The hysteresis width is the
difference between the values of $b_a$ at which the two branches cross $x=1/2$,
obtained by linear interpolation.

\paragraph{Measured rates.} During the measurement window we count the switches $n_{H\to A}$ and $n_{A\to H}$ and the total times $T_H$ and $T_A$ spent by agents in each strategy (the integral over time of the number of agents in the strategy). The per-capita rates are $r_{+}=n_{H\to A}/T_H$ and $r_{-}=n_{A\to H}/T_A$, and $\rho=\eps_{\sw}/(r_{+}+r_{-})$.

\begin{figure}[h]
\centering

\includegraphics[width=0.9\linewidth]{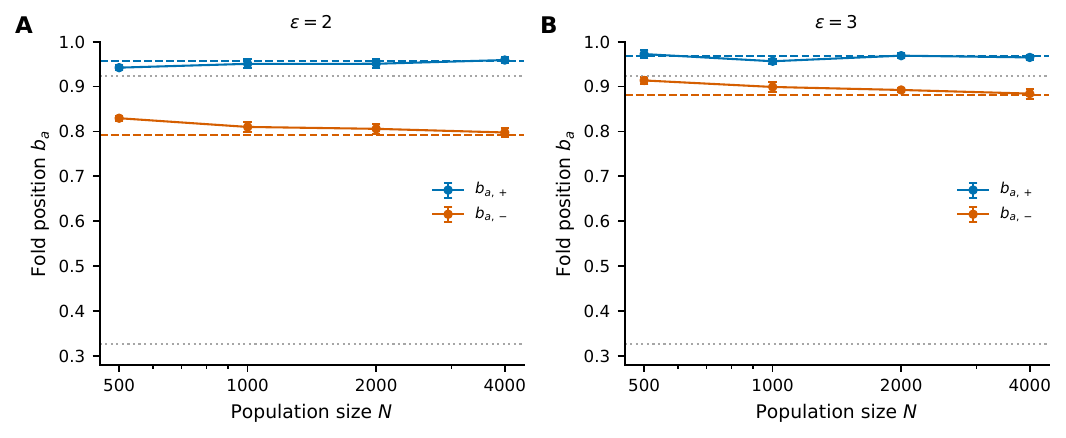}
\caption{\textbf{Finite-size test of the fold positions.} Upper fold $b_{a,+}$ (blue) and lower fold $b_{a,-}$ (orange) measured in the agent-based model as functions of the population size $N$, for $\eps=2$ (left) and $\eps=3$ (right). Dashed lines: folds predicted with $\geff$ ($b_{a,-}\simeq0.791$, $b_{a,+}\simeq0.957$ for $\eps=2$; $b_{a,-}\simeq0.881$, $b_{a,+}\simeq0.967$ for $\eps=3$). Dotted lines: folds of the reduced MFT ($0.325$ and $0.925$). Points: mean and standard deviation over $3$ repetitions, $N=500$, $1000$, $2000$ and $4000$. Other parameters as in \Mfig{fig:agent-hysteresis}.}
\label{supp:fig:SI-finite-size}
\end{figure}

\paragraph{Skill distributions (\Mfig{fig:agent_capacity_distributions}).} Runs start from $x=0.4$ with random strategies and $s_i=1$; after a burn-in, histograms of $s_i$ are accumulated from snapshots every $5$ sweeps.

\subsection{Finite-size test}

In a finite population, fluctuations can carry the state across the barrier separating the two basins before a fold is reached, which narrows the observed hysteresis loop. To measure the folds, we run local quasi-static sweeps in steps of $\Delta b_a=0.0125$ over an interval extending $0.15$ inside and $0.05$ beyond each predicted fold. The upper fold is approached from a mostly autonomous population ($x=0.01$, $s_i=1$) and the lower fold from a mostly AI-reliant, deskilled population ($x=0.99$, $s_i=\eta$ for AI users). Each fold is located where the branch crosses $x=1/2$. Fig.~\ref{supp:fig:SI-finite-size} shows that the measured folds approach the MFT with covariance closure predictions as $N$ grows, while remaining far from the folds of the reduced MFT.

\subsection{Figures parameters}

The full list of parameters is reported in Table \ref{supp:tab:params}.

\begin{table}[h]
\centering
\caption{Parameters of the figures. $\eps$ and $m$ are in rescaled units.
A dash indicates a parameter that does not enter the plotted calculation.}
\label{supp:tab:params}
\small
\begin{tabular}{@{}lccccccccc@{}}
\toprule
Figure & $b_h$ & $c$ & $\eta$ & $\alpha$ & $m$ & $\beta$ & $N$ & $k$ & $\eps$\\
\midrule
Main Fig.~1 & 1 & 0.25 & 0 & 1 & 0.002 & -- & -- & -- & 0.05\\
Main Fig.~2 & 1 & 0.25 & 0 & 1 & 0.002 & 0.5 & 4000 & 50 & 0.05--6\\
Main Fig.~3 & 1 & 1.2 & 0 & 1 & 0.002 & 1 & 3000 & 50 & 0.03, 0.3, 3\\
Main Fig.~4 & 1 & $\omega d$ & 0 & 1 & 0.002 & 0.5 & 4000 & 50 & 2\\
\midrule

Fig.~\ref{supp:fig:SI-nonlinear-payoff} & 1 & 1, 1.2 & 0 & 2, 1/2
    & 0 & -- & -- & -- & --\\
Fig.~\ref{supp:fig:SI-nonlinear-map} & 1 & 0--2 & 0 & 1/2, 1, 2
    & 0 & -- & -- & -- & --\\
Fig.~\ref{supp:fig:SI-nonlinear-portraits} & 1 & 1, 1.2 & 0 & 2, 1/2
    & 0.002 & -- & -- & -- & 0.05\\
Fig.~\ref{supp:fig:SI-nonlinear-branches} & 1 & 0.25, 1 & 0
    & 1/2, 1, 2, 3 & 0.002 & -- & -- & -- & --\\
Fig.~\ref{supp:fig:SI-nonlinear-cusp} & 1 & 0.25, 1 & 0--1
    & 1/2--3 & $m_c$ & -- & -- & -- & --\\
Fig.~\ref{supp:fig:SI-nonlinear-sampling} & -- & -- & -- & 1/2, 2
    & -- & -- & -- & 3, 10, 50 & --\\
Fig.~\ref{supp:fig:SI-nonlinear-agents} & 1 & 1, 1.2 & 0 & 2, 1/2
    & 0.002 & 0.5 & 20000 & global, 50 & 0.05\\
    Fig.~\ref{supp:fig:SI-finite-size} & 1 & 0.25 & 0 & 1 & 0.002 & 0.5
    & 500--4000 & 50 & 2, 3\\
\bottomrule
\end{tabular}
\end{table}

\section{Placing real settings in $\rho$}
\label{supp:sec:realworld}

An unpractised skill decays as $u(t)=u(0)e^{-\eps t}$ under AI use, so its half-life is $t_{1/2}=\ln2/\eps$. The correlation time of an agent's strategy is $\tau=1/\lambda$ (Section~\ref{supp:sec:rates}). Hence
\begin{equation}
    \rho=\frac\eps\lambda=\ln2\,\frac{\tau}{t_{1/2}}\simeq0.7\,\frac{\tau}{t_{1/2}} .
\end{equation}
Both quantities are measurable: $t_{1/2}$ from the decline of unaided performance during sustained AI use \cite{arthur_jr_factors_1998}, and $\tau$ from the autocorrelation of individual decisions to use AI in usage records. For $t_{1/2}\approx50$ days \cite{krakauer_competitive_2026}, strategy choices revised daily give $\rho\approx0.014$, and decisions that persist for several months give $\rho$ of order one.

\clearpage
\endgroup

\end{document}